\documentclass[prc,aps,float,twocolumn,showpacs,nofootinbib,superscriptaddress]{revtex4}
\usepackage{amsmath}
\usepackage{amssymb}
\usepackage{amsfonts}
\usepackage{graphicx}
\usepackage[usenames,dvipsnames]{color}
\usepackage{afterpage}
\usepackage{mathrsfs}
\usepackage{graphicx,latexsym,epsfig}
\usepackage{amsfonts}
\usepackage{mathrsfs}
\usepackage{color}
\usepackage{bm}
\usepackage[colorlinks,linkcolor=red,anchorcolor=blue,citecolor=blue]{hyperref}

\newcommand{\beq}{\begin{equation}}
\newcommand{\eeq}{\end{equation}}
\newcommand{\beqn}{\begin{eqnarray}}
\newcommand{\eeqn}{\end{eqnarray}}

\newcommand{\br}{{\mathbf{r}}}

\newcommand{\bq}{{\mathbf{q}}}

\newcommand{\ket}[1]{|#1\rangle}

\definecolor{mbscolor}{rgb}{0.31, 0.78, 0.47}

\newcommand{\elemA}[2]{\ensuremath{{}^{#1}}\textrm{#2}}

\begin{document}

\title{Beyond mean-field study of elastic and inelastic
       electron scattering off nuclei}

\author{J. M. Yao\footnote{Present address:  Department of Physics, Tohoku University, Sendai 980-8578, Japan}}
\affiliation{Physique Nucl\'eaire Th\'eorique,
             Universit\'e Libre de Bruxelles, C.P. 229, B-1050 Bruxelles,
             Belgium}
\affiliation{School of Physical Science and Technology, Southwest University, Chongqing, 400715 China}

\author{M. Bender}
\affiliation{Universit\'e de Bordeaux, Centre d'Etudes Nucl\'eaires de
Bordeaux Gradignan, UMR5797, F-33175 Gradignan, France}
\affiliation{CNRS/IN2P3, Centre d'Etudes Nucl\'eaires de Bordeaux
Gradignan, UMR5797, F-33175 Gradignan, France}

\author{P.-H. Heenen}
\affiliation{Physique Nucl\'eaire Th\'eorique,
             Universit\'e Libre de Bruxelles, C.P. 229, B-1050 Bruxelles,
             Belgium}

\date{18 December 2014}

\begin{abstract}
\begin{description}
\item[Background] Electron scattering provides a powerful tool to determine charge distributions and transition densities of nuclei. This tool will soon be available for short-lived neutron-rich nuclei.
\item[Purpose] Beyond mean-field methods have been successfully applied to the study of excitation spectra of nuclei in the whole nuclear chart. These methods permit
to determine energies and transition probabilities starting from an effective in-medium nucleon-nucleon interaction but without other phenomenological ingredients.
 Such a method has recently been extended to calculate the charge density of nuclei deformed at the mean-field level of approximation [J.\ M.\ Yao \textit{et al}., Phys.\ Rev.\ C \textbf{86}, 014310 (2012)].
 The aim of this work is to further extend the method to the determination of transition densities between low-lying excited states.
\item[Method] The starting point of our method is a set of Hartree-Fock-Bogoliubov wave
functions generated with a constraint on the axial quadrupole moment and using
a Skyrme energy density functional. Correlations beyond the mean field are
introduced by projecting mean-field wave functions on angular-momentum and
particle number and by mixing the symmetry restored wave functions.
\item[Results] We give in this paper detailed formulae derived for the calculation of densities and form factors. These formulae
are rather easy to obtain when both initial and final states are $0^+$ states but are far from being trivial when one of the states has a finite $J$-value.
Illustrative applications to $^{24}$Mg and to the even-mass $^{58-68}$Ni have permitted to analyse the main features of our method, in particular the effect of deformation on densities and form factors.
An illustration calculation of both elastic and inelastic scattering form factors is presented.
\item[Conclusions] We present a very general framework to calculate densities of and transition densities between low-lying states that can be applied to any nucleus.
To achieve better agreement with the experimental data will require to improve the energy density functionals that are currently used and also to introduce quasi-particle
excitations in the mean-field wave functions.
\end{description}

\end{abstract}

 \maketitle



\section{Introduction}

Electron scattering off nuclei is a powerful tool for studies of nuclear
structure and spectroscopy~\cite{Hofstadter56,Alder56,Forest66,
Uberall71,Barrett74,Dreher74a,Donnelly75,Friar75,Heisenberg81,Heisenberg83,
Donnelly84,Sick85,Vries87,Frois87,Hodgson92,Walecka04a}.
It allows to determine the
charge distribution of nuclear ground states, as well as of the transition
charge and current densities from the ground state to excited states.
More global properties can be extracted from a detailed knowledge of charge distribution,
like charge radii. Parameters characterizing the
extension and surface thickness of the nuclear density can also be derived~\cite{FV82,Friedrich86}.
From the form factors for inelastic electron scattering at low transferred
momentum $q$, the spin and parity of excited states and the multipole
transition strengths can be determined in a model-independent
manner~\cite{Uberall71,Heisenberg83}. At larger values of $q$, the form
factors present an insight into the spatial location of the transition
process, which cannot be accessed from the integral over this function
provided by the measurement of $B(EL)$ values in Coulomb excitation
or lifetime measurements.
Thereby, electron scattering does not only provide a powerful alternative
to many other types of nuclear structure studies, but also complements them
by giving access to levels and transitions that are undetectable in
photoexcitation and $\gamma$-ray spectroscopy, such as for instance levels
excited by monopole transitions or transitions of high multipolarity.

As all electron-nucleus scattering experiments of the past used fixed
or gas targets, only stable and a very few long-lived nuclides could
be studied so far.
This will change with the set-up of electron-RIB collider experiments.
The SCRIT (Self Confining Radioactive Isotope Target)
project~\cite{Wakasugi04,Suda05,Suda09} is under construction at RIKEN (Japan)
and the ELISe (ELectron-Ion Scattering in a storage ring) project is planned
for FAIR (Germany)~\cite{Simon07a,Antonov11}. When being realised, the
charge densities and transition charge densities of short-lived nuclides,
in particular neutron-rich nuclei, will be measured at both installations.

Data from electron scattering are often interpreted in terms of parameterized
macroscopic density and transition density distributions, such as the ones
of Helm~\cite{Helm56}, Tassie~\cite{Tassie56} or Friedrich
\textit{et al}.~\cite{FV82,Friedrich86}. They all have in common that
some functional form of the ground-state or transition charge densities
is postulated and its parameters adjusted  to reproduce the data. Such analysis
provides an insight into the gross features of the ground state and
transition charge density distribution and the resolution of their
details \cite{Dreher74a}. For a more detailed analysis, however, it is desirable to calculate the
form factors from the same microscopic models that are also used to
describe nuclear structure and spectroscopy. Most of them have been
used to describe one and/or the other in the past.

\begin{itemize}
\item
Shell model calculations in small valence spaces have been used
to calculate transition densities between states in light nuclei
\cite{Brown83}. Some heavier nuclei have been calculated within the framework of the
interacting Boson approximation \cite{Dieperink78a}. In both cases,
the truncation of the model space requires to introduce effective
charges and/or even explicitly calculated core polarization effects
\cite{Brown83,Horikawa77,Sagawa87,Yokoyama89,Radhi03}.
The no-core shell model, available only for light nuclei,
is better suited in that respect \cite{Karataglidis07,Radhi08a}.

\item
Methods based on self-consistent mean fields \cite{Bender03} are
a natural choice for such calculations, in particular for heavy nuclei,
as they use a model space that comprises all occupied single-particle
levels and an effective interaction or energy density functional (EDF)
that is designed to reproduce nuclear saturation.
Indeed, electron scattering form factors of spherical nuclei have already
been studied in the pioneering papers of this field
\cite{Negele70,Miller72,Vautherin72,Decharge80}.
More recent studies emphasize the possible isospin dependence of charge
form factors of spherical nuclei \cite{Wang04,Antonov05,Roca08a,Roca12a}.
With the exception of excitation to collective rotational states in
well-deformed nuclei \cite{Negele77,Moya80a,Berdichevsky88,Sarriguren89}, pure mean-field calculations,
however, are limited to ground-state densities. They also miss
correlations from fluctuations in collective degrees of freedom
and from symmetry restoration that should be considered for
non-spherical nuclei.

\item The random phase approximation (RPA) (or the quasi-particle RPA) on top of mean-field calculations has been applied to
spherical nuclei to study the ground state and transition charge densities~\cite{Faessler76,Reinhard79,Gogny79,Decharge83,Esbensen83,Barranco87,Johnson88,Sil08,Nobre11}.
    The extension of this framework to the density and transition density for deformed nuclei is, however, not trivial.

\item There also has been a number of studies where various electric and
magnetic electron scattering
form factors of deformed nuclei have been calculated by angular momentum
projection of mean-field wave functions. To limit the computational cost, the
wave functions were either restricted to be of a simple form~\cite{Abgrall74a},
or the symmetry restoration was approximated in one way or the
other~\cite{Zaringhalam77,Moya78a,Guerra80,Moya80a,Dieperink87,Graca88,Berdichevsky88,Nishimura88}. For a presentation of the main aspects of these developments, see the review~\cite{Guerra86}.
\end{itemize}

Recently, we have used the framework of the particle-number and
angular-momentum projected generator coordinate method (GCM) based on axial Hartree-Fock-Bogoliubov (HFB) states and a
non-relativistic Skyrme energy density functional to calculate
the ground state density of even-even nuclei~\cite{Yao12}, demonstrating how the
correlations brought by going beyond a mean field approach can quantitatively,
even qualitatively alter the density profile predicted by pure
mean-field methods. The same technique has been subsequently implemented
in the relativistic framework using covariant energy density functionals~\cite{Yao13a,Wu14a,Mei14}.
Here, we extend the formalism of Ref.~\cite{Yao12} to transition densities
between low-lying excited states and the corresponding form factors as
accessible by electron scattering. The emphasis of this first exploratory study is on the impact of static
and dynamic quadrupole deformations on the transition density between
low-lying collective states. Similar developments based on an angular-momentum and parity projected
GCM with (non-paired) HF states, also using Skyrme interactions
have been recently reported in Ref.~\cite{Fuk13}, but limited to the simple case of
elastic and inelastic transitions between $0^+$ states.

The paper is organized as follows. In Sec.~\ref{Sec.II} we
present the relevant formulae for the description of electron scattering
off nuclei and the formalism for the calculations of nuclear density
distribution and transition density for low-lying states in the framework
of projected GCM based on axially deformed HFB states.
In Sec.~\ref{Sec.III}, we present an illustrative calculation of both
elastic and inelastic scattering form factors for $^{24}$Mg.
Section~\ref{Sec.IV} details an application to the transition densities
in even-mass $^{58-68}$Ni. The static and dynamic deformation effects
on nuclear charge densities, transition charge densities and form factors
will be discussed in detail.
Section~\ref{Sec.V} summarizes our findings, and four appendices provide
further technical details on the calculation of nuclear form factors and
the transition density.

\section{Formalism}
\label{Sec.II}

\subsection{Beyond mean-field description of nuclear states}

Our beyond-mean-field method restores two of the symmetries relevant
for nuclear spectroscopy that are broken by the self-consistent
mean field HFB method by projection on particle number and angular
momentum. Fluctuations in shape degrees of freedom are described
by the superposition of projected HFB states with different
intrinsic deformations. The same formalism that is used to calculate
operator matrix elements between projected states can be used to
calculate projected densities and their form factors. Before entering
into the details of their calculation, we first recall the main
features of the method.

\subsubsection{Quadrupole deformed HFB states}

A set of deformed HFB states is generated by solving the HFB equations
including a constraint on the axial quadrupole moment using an updated
version of the code first described in \cite{Ter96a}.
The states are restricted to be time-reversal invariant and reflection
symmetric, which implies that they are eigenstates of parity with
eigenvalue $+1$. The HFB equations are complemented by the Lipkin-Nogami
prescription to avoid the unphysical breakdown of pairing correlations
at low density of single-particle levels around the Fermi energy.

The single-particle wave functions are discretized on a
three-dimensional Cartesian coordinate-space mesh~\cite{Bonche05}. The
step size of $0.8 \, \text{fm}$ ensures a good accuracy in the solution of
the mean-field equations.

Throughout this study, we use the Skyrme parametrization SLy4
\cite{Chabanat98} together with a pairing energy functional of
surface character~\cite{Rigollet99} with parameters
$\rho_0 = 0.16 \, \text{fm}^{-3}$ for the switching density and
$V_{0} = -1000 \, \text{MeV} \, \text{fm}^3$ for the pairing strength.
A soft cutoff at $\pm 5$ MeV around the Fermi energy is
used when solving the HFB equations as described in Ref.~\cite{Rigollet99}.

\subsubsection{Projected GCM states}

The GCM wave function~\cite{Ring80} is constructed as a superposition
of both particle-number and angular-momentum projected HFB wave functions corresponding to different deformations
$| q \rangle$
\begin{equation}
\label{eq_GCM:10}
\ket{J M \mu}
= \sum_{q} F^{J}_{\mu, q} \hat{P}^J_{M0} \hat{P}^{N} \, \hat{P}^{Z} \, \ket{q} \, ,
\end{equation}
where $\mu$ labels different collective states for a given angular
momentum $J$. This \textit{ansatz} can cover a wide variety of
situations, such as small fluctuations around a spherical
or well-deformed minimum of a deep and steep potential well, wide
fluctuations in soft nuclei, or mixing of states in different minima
of the energy surface.

The operators $\hat{P}^{Z}$ and $\hat{P}^{N}$ project on proton and neutron number,
\begin{equation}
\label{proj_N}
\hat{P}^N
= \frac{1}{2\pi} \int^{2\pi}_0 \!  d\varphi \, e^{\text{i}\varphi(\hat N-N)} \, ,
\end{equation}

and $\hat{P}^J_{MK}$ extracts eigenstates of total angular momentum $J$ with $z$ component $M$
\beqn
\label{proj_J}
\hat{P}^J_{MK}
&=& \frac{\hat{J}^2}{8\pi^2}
  \int \! d\Omega \; \mathcal{D}^{J*}_{MK} (\Omega) \, \hat{R}(\Omega) \, ,
\eeqn
where $\hat{J}^2\equiv2J+1$, $\hat{R}(\alpha,\beta,\gamma) \equiv
e^{-i\alpha \hat{J}_x} \, e^{-i\beta \hat{J}_y} \, e^{-i\gamma \hat{J}_z}$ is the rotation operator and $\mathcal{D}^{J}_{MK} (\alpha,\beta,\gamma)$ the Wigner $D$-function. Both depend on the Euler angles, for which we will use the shorthand notation
$\Omega \equiv (\alpha, \beta, \gamma)$ whenever possible. The volume element of the integration over Euler angles is given by $d\Omega \equiv d\alpha \, d\beta \, \sin(\beta) \, d\gamma$.  Only a $K=0$ component can be picked by $\hat{P}^J_{MK}$ from an HFB state that is axially symmetric around the $z$ axis. Therefore, the index $K$ will be dropped for simplicity.

The weight factors $F^{J}_{\mu, q}$ and the energies
of the states $\ket{J M \mu}$ are obtained by solving a
Hill-Wheeler-Griffin equation~\cite{Ring80}
\begin{equation}
\label{eq_GCM:20}
\sum_{q} \Big( \mathcal{H}^{J}_{q' q} - E_\mu^{J}\mathcal{N}^{J}_{q' q}
          \Big) \, F_{\mu,q}^{J} = 0
\, ,
\end{equation}
for each value of $J$, where the norm kernel
$\mathcal{N}^{J}_{q' q} = \langle q'\vert \hat{P}^J_{00} \hat{P}^{N} \, \hat{P}^{Z}\vert q \rangle$ and the energy
kernel $\mathcal{H}^{J}_{q' q}$ is a functional of mixed densities~\cite{Lacroix09}. More details about the calculations can be
found in Ref.~\cite{Bender08} and references given therein.

As the projected mean-field states do not form an orthogonal
basis and the weights $F^{J}_{\mu, q}$ in Eq.~(\ref{eq_GCM:10}) are not
orthogonal functions, a set of orthonormal collective wave functions
$g^{J}_{\mu, q}$ is constructed as~\cite{Ring80}
\begin{equation}
\label{eq_GCM:30}
g^{J}_{\mu, q}
= \sum_{q'} \big( \mathcal{N}^{J} \big)^{1/2}_{q q'} \, F^{J}_{\mu, q'}
\, ,
\end{equation}
but the modulus square of $g^{J}_{\mu, q}$ does not represent the probability to find the
deformation $q$ in a GCM state $\ket{J M \mu}$. In a GCM based on
axial states, however, the $g^{J}_{\mu, q}$ do nevertheless provide a good
indication about the dominant configurations in the collective
states $\ket{J M \mu}$.

\subsection{Form factors in electron scattering}

\subsubsection{General framework}

Our aim is to show how to calculate form factors and transition densities
in the framework of our model. We will therefore not enter into the
details of the process of scattering electrons off nuclei itself
and limit the presentation to those elements of the formalism that are
necessary to compute densities, transition densities and their form
factors in a form that can then be compared to experiment.

We use the framework of the plane-wave Born approximation (PWBA).
The incident and outgoing electrons are described by plane waves
$e^{i\mathbf{k}_i\cdot\br}$ and $e^{i\mathbf{k}_f\cdot\br}$ with momenta
$\mathbf{k}_i$ and $\mathbf{k}_f$ and energies $E^e_i$ and $E^e_f$,
respectively. The differential cross section for electron scattering from a spin-less
nucleus is given by~\cite{Alder56,Forest66,Uberall71,Donnelly75}
\beq
\label{cross-section2}
\dfrac{d\sigma}{d\Omega}
= \dfrac{d\sigma_{\rm M}}{d\Omega} \sum_{L \ge 0} | F_L(q) |^2 \, ,
\eeq
 which is the product of the Mott cross section $d\sigma_{\rm M}/d\Omega$
describing the cross section for scattering off a point-like target with charge
$Z$~\cite{Hofstadter56,Heisenberg82} times the sum of form factors $F_L(q)$
that represent its modification by the nucleus having a finite size and an
 internal structure. The cross section depends on the momentum transfer
$q = |\mathbf{k}_f-\mathbf{k}_i |\simeq 2 \sqrt{E^e_iE^e_f} \,
\sin (\theta /2)$, where ${k}_i(E^e_i)$ and ${k}_f (E^e_f)$  are the momenta (energies) of the
incoming and outgoing electron, and $\theta$ the angle between
$\mathbf{k}_i$ and $\mathbf{k}_f$.

The longitudinal Coulomb (CL) form factor $F_L (q)$ in (\ref{cross-section2}) for an angular momentum
transfer $L$ is the Fourier-Bessel transform of the transition density
$\rho^{J_f \mu_f}_{J_i \mu_i,L}(r)$ from an initial state
$| J_i M_i \mu_i \rangle$ to a final nuclear state $| J_f M_f \mu_f \rangle$
 \beq
 \label{FF1}
 F_L (q)
 = \dfrac{\sqrt{4\pi}}{Z}
   \int^\infty_0 \! dr \; r^2 \, \rho^{J_f \mu_f}_{J_i \mu_i,L}(r) \,
   j_L (qr) \, ,
 \eeq
where the coefficient $\sqrt{4\pi}/Z$ is chosen so that the elastic
part ($J_f=J_i, \mu_f=\mu_i$) of the form factor $F_0(q)$ is unity at $q=0$. In this expression, $\rho^{J_f \mu_f}_{J_i \mu_i,L}(r)$ is the reduced transition density that will be related to GCM matrix elements in the next section.

In electron scattering off nuclei, the Coulomb attraction accelerates
the electrons when they approach the nucleus and the electron
wave is focused onto the nucleus. As a consequence, an experiment
actually samples the form factor at a larger momentum transfer than
given by the asymptotic values of the kinematic variables.
This can be corrected for by plotting the experimental data measured for
a given $q$~\cite{Uberall71,Brown83,Heisenberg83} as a function of the corresponding
``effective" momentum transfer $q_{\text{eff}}$
 \beq
 q_{\rm eff}
 = q \left( 1 + \dfrac{3Ze^2}{2E^e_{i} R_{\rm ch}} \right) \, ,
 \eeq
where $R_{\text{ch}}$ is the \textit{equivalent hard sphere radius} of the
nucleus that is related to its rms charge radius $r_{\text{ch}}$
by $R_{\text{ch}} = \sqrt{5/3} \, r_{\text{ch}}$. Values for $r_{\rm ch}$ used
in what follows are taken from a compilation of experimental data~\cite{Angeli04}.
It was concluded in Ref.~\cite{Brown83} that the Coulomb distortion effect
of the scattered electrons is mostly taken into account by this prescription
and that there is no significant advantage to replacing PWBA calculations for
inelastic scattering with more involved distorted-wave Born approximation
(DWBA) calculations, in particular when considering the limitations
in precision of both data and their theoretical modeling.

A correction for the finite size of the proton is introduced by folding
all calculated point proton densities with a Gaussian form
factor~\cite{Negele70}, for example
 \beq
 \label{charge-folding}
 \rho_{\rm ch} (\br)
  = \left(\dfrac{1}{a \sqrt{\pi}}\right)^{3}
    \int \! d^3r' \; \exp \left[-\dfrac{(\br-\br')^2}{a^2} \right] \,
    \rho_p (\br') \, ,
 \eeq
where $a = \sqrt{2/3} \, \langle r^2 \rangle^{1/2}_p = 0.65 \, \text{fm}$. When high
precision is required, more detailed parametrizations of the proton and
neutron charge distributions have to be used together with relativistic
corrections, cf.\ \cite{Brown83,Bender03} and references therein.

A correction for the spurious center-of-mass (COM) motion related to the breaking
of translational invariance by the nuclear mean field should also be
introduced. A rigorous way to remove it is to project on the COM, which,
however, is difficult to achieve in combination with angular-momentum
projection for deformed states. As has been shown in such calculations
for spherical mean-field states \cite{Schmid90a,Schmid91a,Rodriguez04a},
the relative importance of the c.m.\ correction quickly fades away for
heavy nuclei. A more economical approximation still in use \cite{Fuk13}
is the harmonic oscillator approximation first proposed in Ref.~\cite{Tassie58},
where the calculated charge form factor is corrected by folding it with
a COM motion correction $F_{\text{ch,corr}}(q) = F_{\text{ch}}(q) \, G_{\rm cm}(q)$
obtained in harmonic oscillator approximation
 \beq
  \label{CM}
 G_{\rm cm}(q) = \exp \big[ q^2b^2/(4A) \big] \, ,
 \eeq
where $A=N+Z$ and $b$ being a suitable oscillator length parameter
\cite{Brown83}. In what follows, we will use $b=\sqrt{\hbar/m\omega_0}$, where $m$ is the bare nucleon mass and the frequency $\omega_0$ is given by $\hbar\omega_0=41 \, A^{-1/3} \; \text{MeV}$.
As we will show below in Fig.~\ref{Mg24:F2}, already
for \elemA{24}{Mg} the effect of the COM motion correction is too
small to be relevant for the purpose of our discussion.

\subsubsection{Transition density between GCM states}

To calculate form factors (\ref{FF1}) for elastic and inelastic electron
scattering and transition matrix elements, we need to determine the {\em reduced transition density}  $\rho^{J_f \mu_f}_{J_i \mu_i,L}(r)$ as a function of the radial coordinate $r$.
 We now derive its relation to the {\em 3D transition density}   $\rho^{\alpha_f}_{\alpha_i}(\br)$ between the initial $| \alpha_i \rangle$ and a final   $| \alpha_f \rangle$  GCM states
 \beqn
  \label{tdens:gcm}
  \rho^{\alpha_f}_{\alpha_i}(\br)
  &\equiv& \langle \alpha_f | \hat \rho(\br) | \alpha_i \rangle
           \nonumber\\
  & =    & \sum_{q'q} F_{\mu_f, q'}^{J_f*}
           F_{\mu_i, q}^{J_i}
           \rho^{\sigma_fq'}_{\sigma_iq} (\br) \, ,
 \eeqn
where we have introduced the shorthand notations  $\alpha\equiv\{JM\mu\}$ and $\sigma\equiv\{JM K\}$.
With the exception of the appendices, we restrict the discussion to axial states and $\sigma\equiv\{JM 0\}$.
The density operator is defined as
$\hat{\rho}(\br) \equiv \sum_i \delta(\br-\br_i)$, where $\br$ is the
position at which the transition density is calculated, and $\br_i$
the position of the $i$-th nucleon.

The kernel of the {\em 3D transition density}  between two axial HFB states projected on particle numbers $N, Z$ and angular momentum $J$ is determined by
 \beq
 \label{tdens:kernel0}
  \rho^{\sigma_fq'}_{\sigma_iq} (\br)
 \equiv \langle q' | \hat{P}^{J_f}_{0 M_f} \, \hat{\rho}(\br) \,
      \hat{P}^{J_i}_{M_i0} \, \hat{P}^N \, \hat{P}^Z | q \rangle \, .
 \eeq
The calculation of a matrix element like Eq.~\eqref{tdens:kernel0} can be simplified for an operator that is a
spherical tensor by eliminating one of the two rotations~\cite{Bender08,Yao10,Rodriguez10}. The density operator, however, is not
a spherical tensor operator, the evaluation
of its matrix elements is considerably more complicated as  both rotations in
Eq.~\eqref{tdens:kernel0} will have to be carried out numerically.

Inserting the explicit expressions for the projection operators into
Eq.~\eqref{tdens:kernel0}, one obtains for the transition density kernel (see Appendix~\ref{append2} for further details),
 \beqn
 \label{tdens:kernel}
\rho^{\sigma_fq'}_{\sigma_iq} (\br)
  &=&
      \dfrac{\hat{J}^2_f}{8\pi^2} \int \! d\Omega^\prime \;
       D^{J_f}_{M_f0}(\Omega^\prime) \nonumber\\
  & &\times \sum_{K} D^{J_i\ast}_{M_iK}(\Omega^\prime) \,
      \hat{R}(\Omega^\prime) \, \rho^{J_i K0}_{q'q}(\br) \, ,
 \eeqn
where $\rho^{J_i K0}_{q'q}(\br)$  for axially deformed nuclei is simplified as
 \beqn
 \label{rho_JKNZ}
 \rho^{J_i K0}_{q'q}(\br)
 & \equiv & \dfrac{\hat J^2_i}{2} \int^\pi_0 \! d\beta \, \sin(\beta) \,
          d^{J_i}_{K 0}(\beta) \nonumber\\
 &&\times
 \langle q' | \hat{\rho}(\br) \,
 \hat P^{N} \hat P^{Z} \hat{R}_y(\beta) | q \rangle \, ,
 \eeqn
The calculation of the density \eqref{rho_JKNZ} requires the determination
of non-diagonal matrix elements of the density operator between a rotated
and a non-rotated state analogous to the calculation of projected matrix
elements of tensor operators~\cite{Bender08,Yao10,Rodriguez10}.
As shown in Ref.~\cite{Yao10}, when $x$-signature is preserved,
the integrant $\rho_{q'q}(\br,\beta)
\equiv \langle q' | \hat{\rho}(\br) \, \hat P^{N} \hat P^{Z}
\hat R_y(\beta) | q \rangle$ presents a symmetry in $\beta$ with respect
to $\pi/2$
 \beq
 \rho_{q'q}(x,y,z,\pi-\beta) = \rho_{q'q}(-x,y,z,\beta) \, ,
 \eeq
which can be used to reduce the number of density overlaps to be
calculated explicitly by a factor of two.

Compared to the calculation of operator matrix elements, the unfamiliar
element in the calculation of the projected transition
density kernels \eqref{tdens:kernel} is that the integration
over $\Omega'$ cannot be carried out analytically. Instead,
Eq.~\eqref{tdens:kernel} involves the rotation of the
density $\rho^{J_i K0}_{q'q}(\br)$ as a whole.

In a 3D coordinate space representation as used here, a rotation requires
an interpolation of the rotated function, as the rotated coordinates of
the mesh points do in general not fall back on the mesh. In our case,
the integration over $d \cos(\beta)$ is discretized using a Gauss-Legendre
quadrature with 24 points in the interval $[-1,+1]$, which is sufficient
for the low values of $J$ considered here. The corresponding rotations
$\hat{R}_y(\beta)$ in \eqref{rho_JKNZ} are carried out with the same
accurate Lagrange-mesh technique~\cite{Baye86,Valor00} that is also used
to evaluate operator matrix elements in our codes.

To perform the rotation of $\rho^{J_i K 0}_{q'q}(\br)$ in Eq.~\eqref{tdens:kernel},
it turned out that, instead of a rotation of the density
followed by an integration over Euler angles, it is advantageous
to expand $\rho^{J_i K0}_{q'q}(\br)$ into spherical harmonics first.
Using the transformation of spherical harmonics under rotation and
some further angular-momentum algebra that is detailed in
Appendix~\ref{append3}, the integrals over Euler angles $\Omega'$ in
Eq.~\eqref{tdens:kernel} can be transformed into integrals over
spatial angles that are much easier to carry out
 \beqn
\rho^{\sigma_fq'}_{\sigma_iq} (\br)
& = & \dfrac{\hat J^2_f}{\hat J^2_i}\sum_{K\lambda \nu^\prime}
      \langle J_f0 \lambda K | J_iK\rangle \,
      \langle J_f M_f \lambda \nu^\prime | J_i M_i \rangle \nonumber\\
 && \times  \rho^{J_iK0}_{q'q;\lambda K} (r) \,
       Y_{\lambda\nu^\prime}(\hat \br) \, ,
\eeqn
where $\rho^{J_iK0}_{q'q;\lambda K} (r)$ is given by
 \beq
 \rho^{JK0}_{q'q;\lambda K} (r)
 = \int \! d\hat\br' \; \rho^{J_iK0}_{q'q}(r, \hat \br') \,
    Y^\ast_{\lambda K}(\hat \br') \, .
 \eeq
Finally, the 3D transition density of an axially deformed nucleus is given by
\beqn
  \label{3Dtdens:gcm}
  \rho^{\alpha_f}_{\alpha_i}(\br)
  & = &
          \dfrac{\hat J^2_f}{\hat J^2_i}\sum_{K\lambda \nu^\prime}
      \langle J_f0 \lambda K | J_iK\rangle
      \langle  J_f M_f \lambda \nu^\prime | J_i M_i \rangle
       \nonumber\\
 &&\times Y_{\lambda\nu^\prime}(\hat \br) \int   d\hat\br'  \rho^{J_fJ_iK0}_{\mu_f\mu_i}(r, \hat \br')
    Y^\ast_{\lambda K}(\hat \br') \, ,
 \eeqn
where we have introduced a configuration-mixing \textit{pseudo GCM density}\footnote{This \textit{pseudo GCM density} summarizes all the information related to the GCM calculation but it is not an observable. }
$\rho^{J_fJ_iK0}_{\mu_f\mu_i}(\br)$
\beq
\label{pseudo_density}
 \rho^{J_fJ_iK0}_{\mu_f\mu_i}(\br)
 \equiv
 \sum_{q'q} F^{J_f0\ast}_{\mu_f,q'} \, F^{J_i0}_{\mu_i,q} \,
 \rho^{J_iK0}_{q'q}(\br) \, .
\eeq
After some further algebraic manipulations, one obtains the expression of  the radial part of the 3D transition density, namely the \textit{reduced transition density} $\rho^{J_f \mu_f}_{J_i \mu_i,L}(r)$, cf.~(\ref{reducedTD_axial})
  \beqn
 \label{trans-r}
 \rho^{J_f \mu_f}_{J_i \mu_i,L}(r)
&=& (-1)^{J_i-J_f} \dfrac{\hat J^2_f}{\hat J^2_i}
      \sum_{K} \langle J_f0 L K | J_i K\rangle \nonumber\\
 && \times     \int \! d\hat\br \; \rho^{J_fJ_iK0}_{\mu_f\mu_i}(\br) \,
    Y^\ast_{LK}(\hat \br)
\eeqn
that is experimentally accessible via electron scattering.

Compared to the direct evaluation of Eq.~\eqref{tdens:kernel},
the expansion in spherical harmonics has the practical advantage to separate the radial dependence of $\rho^{J_f, \mu_f}_{J_i \mu_i,L}(\br)$, which is specific to each state, from its angular dependence that is completely determined
by the angular momentum quantum numbers of the states.

The integration  over the angular part
of $\br$ in Eq.~\eqref{trans-r}
is discretized using a Gauss-Legendre quadrature with 20~points for
the cosine of the polar angle $\cos(\theta)$ and a trapezoidal rule
with 20~points for the azimuthal angle $\varphi$. To carry out the
integral, the density $\rho^{J_iK0}_{q'q}(\br)$ that is calculated
on a equidistant Cartesian mesh has to be interpolated to the mesh
points in spherical coordinates by using the Lagrange-mesh interpolation
\cite{Baye86}.
The step size $dx$ of the original Cartesian mesh is kept for the radial
coordinate~$r$.

\subsubsection{Transition densities in some special cases}

The expression for the inelastic scattering transition density (TD), given by Eq.~\eqref{3Dtdens:gcm} simplifies greatly
if the initial state is a $0^+$ state
 \beqn
 \label{gstoes}
 \rho^{\alpha_f}_{0^+_{\mu_i}}(\br)
 &=&  Y^\ast_{J_fM_f}(\hat{\br}) \int \! d\hat{\br}^\prime \; \rho^{J_f000}_{\mu_f\mu_i}(r, \hat \br') \,
     Y_{J_f0}(\hat \br^\prime) \, .
 \eeqn
As expected, the angular part of this TD is given by $Y^\ast_{J_fM_f}(\hat \br)$. The reduced transition density becomes
  \beqn
 \rho^{J_f \mu_f}_{0 \mu_i,L}(r)
&=& \hat J_f  \int \! d\hat\br \; \rho^{J_f000}_{\mu_f\mu_i}(\br) \,
    Y_{J_f0}(\hat \br) \, \delta_{J_f L} \, .
\eeqn
For a well-deformed nucleus, that
can be described by a single axial HFB configuration $| q_0\rangle$ and assuming that the overlap between
the rotated wave function and the original one can be approximated
by a $\delta(q-q_0)$ function, the pseudo GCM density $\rho^{J_fJ_iK0}_{\mu_f\mu_i}(\br)$
reduces to the intrinsic density, projected on particle numbers, $\rho^{NZ}_{q_0q_0}(\br) \equiv \langle q_0 | \hat{\rho}(\br) \, \hat{P}^N \, \hat{P}^Z | q_0 \rangle$.
The transition density $\rho^{L\mu_f}_{01,L}(r)$ in Eq.~\eqref{trans-r} is then simply given by
 \beq
 \label{Rotional_model}
 \rho^{L\mu_f}_{01,L}(r)
= \hat{L} \int \! d\hat{\br} \;  \rho^{NZ}_{q_0q_0}(\br) \, Y_{L0}(\hat \br) \, ,
 \eeq
showing that we recover the rigid-rotor model for well-deformed nuclei. The quality of this approximation is quickly
deteriorating with increasing $L$-values, as illustrated in Refs.~\cite{Zaringhalam77,Heisenberg83}.

Putting $\alpha_i=\alpha_f=\alpha$ in Eq.~\eqref{3Dtdens:gcm}, the 3D density for the GCM state $|\alpha\rangle$ is given by
\beqn
 \label{density:JM}
  \rho^{\alpha}_{\alpha}(\br)
  & = & \sum_{\lambda} Y_{\lambda0}(\hat{\br}) \,
      \langle  J  M \lambda 0 | J  M \rangle
     \sum_K \langle J 0 \lambda K | J K \rangle
       \nonumber\\
 && \times \int \!  d\hat{\br}' \; \rho^{JJ K0}_{\alpha \alpha}(r, \hat{\br}')
    \, Y^\ast_{\lambda K}(\hat \br') \, .
 \eeqn
For the ground state $0^+_1$, it is just the average of the pseudo GCM density $\rho^{0000}_{11}(r, \hat \br')$ over the angular coordinates
\begin{eqnarray}
\label{eq_GCM:70}
  \rho^{0^+_1}_{0^+_1}(\br)
  &=& \dfrac{Y_{00}(\hat \br)}{\sqrt{4\pi}}
      \int \!  d\hat{\br}' \; \rho^{0000}_{11}(r, \hat \br'), 
\end{eqnarray}
which obviously is spherically symmetric. This density has been
recently determined for various light systems using the symmetry-restored
GCM method \cite{Yao12,Yao13a,Wu14a}.

\subsubsection{Multipole transition matrix elements}

The multipole transition matrix elements that are frequently calculated
in angular-momentum projected GCM calculations are related to the transition density~\eqref{trans-r} through
 \beqn
 \label{TME}
 M^{J_f \mu_f}_{J_i \mu_i,L}
 &=& \int^\infty_0 \! dr \, r^{L+2} \, \rho^{J_f \mu_f}_{J_i \mu_i,L}(r)
     \nonumber\\
 &=& \hat{J}^{-1}_i \sum_{q'q}
     F_{\mu_f,q_f}^{J_f*} F_{\mu_i,q_i}^{J_i}
 \langle J_f q' || \hat Q_{L} || J_i q \rangle \, ,
 \eeqn
where the reduced matrix element of the multipole operator
$\hat Q_{LM}(\br) \equiv r^L \, Y_{LM} (\hat{\br})$ is defined by
 \beqn
 \langle J_f q^\prime || \hat Q_{L}(\br) || J_i q \rangle
 &\equiv &
 \dfrac{\hat J^2_f\hat J^2_i}{2} (-1)^{J_f}\sum_{M}
 \begin{pmatrix}
 J_f & L &  J_i \\
  0  & M &  -M
 \end{pmatrix}  \nonumber\\
 & & \times \int^\pi_0 \! d\beta \, \sin(\beta) \,  d^{J_i\ast}_{-M0}(\beta)
      \nonumber\\
 & & \times \langle q^\prime | r^L \, Y_{LM} (\hat{\br}) \,
      \hat{P}_{N} \hat{P}_{Z} \hat{R}_y (\beta) | q \rangle \, .
 \nonumber \\
 \eeqn
The electric multipole transition strengths $B(EL: \alpha_i\to \alpha_f)$ are then given by the square of the proton part of the transition matrix element $ M^{J_f \mu_f}_{J_i \mu_i,L}$ (abbreviated with $M^p_L$). More details will be given in Appendix~\ref{append4}.

There have been efforts to deduce the multipole transition matrix elements
$M^p_L$ and $M^n_L$ of protons and neutrons by combining Coulomb excitation
and ($p, p^\prime$) measurements~\cite{Terrien73}, which, however, requires
model assumptions at several stages of the analysis. While their
experimental determination remains debatable, it turns out that the
comparison between the calculated $M^p_L$ and $M^n_L$ sheds light on the
relative contributions by the neutrons and protons to the nuclear
excitation, and therefore provide an insight into the isospin nature
of the \textit{calculated} excitation modes. The deviation of a factor $\eta$
defined as
 \beq
 \label{eta}
 \eta = \dfrac{M^n_L/M^p_L}{N/Z}
 \eeq
from 1.0 is then interpreted as the measure of the isovector character of
the excitation~\cite{Terrien73}. This quantity provides a tool to study
the isospin nature of the excitations, as the multipole
moments of neutrons can be easily calculated in the same way as the
ones of protons.

\section{Illustrative application to $^{24}$M\lowercase{g}}
\label{Sec.III}

\begin{figure}[ht!]
\begin{center}
\includegraphics[clip=,width=0.40\textwidth]{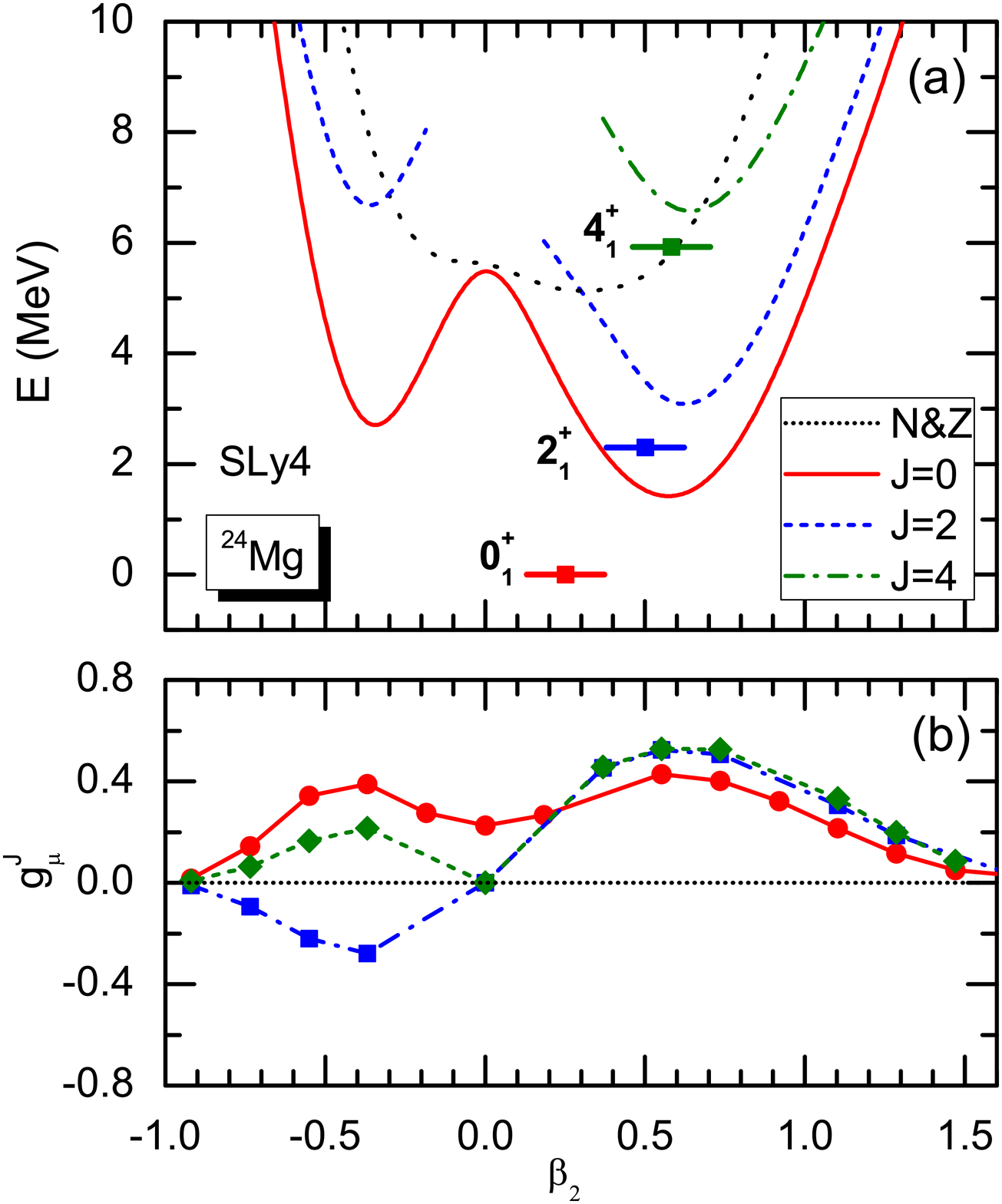}
\end{center}\vspace{-0.5cm}
\caption{\label{Mg24:pec}%
(color online) (a) Total energy (normalized to the $0^+_1$ state) for
the particle-number-projected HFB states (N\&Z) and for the particle-number
and angular-momentum projected states (curves for $J=0$, 2, and 4)
for $\elemA{24}{Mg}$ as a function of the intrinsic mass quadrupole
deformation of the mean-field states. The solid square dots indicate
the lowest GCM solutions, which are plotted at their average deformation
$\bar\beta_{J\mu}$.
(b) Collective wave functions $g_{\mu,q}^{J}$ (cf.\ Eq.~(\ref{eq_GCM:30}))
of the $0^+_1$, $2^+_1$, and $4^+_1$ states.}
\end{figure}

The nucleus $\elemA{24}{Mg}$ has been used as a testing ground for many
implementations of beyond-mean-field models
\cite{Valor00,Guzman02,Niksic06,Bender08,Yao10,Rodriguez10}.
The results presented here are an extension of previous studies. The mass quadrupole moment is discretized with a
step size $\Delta q=40 \, \text{fm}^2$, ranging from $-200 \, \text{fm}^2$
to $+360 \, \text{fm}^2$. This choice ensures a good convergence of the
GCM calculation. The excitation spectra obtained here are the
same as those reported for axial calculations in Ref.~\cite{Bender08}.

 The energy curves obtained after projection on particle numbers only
and after simultaneous projections on particle numbers and angular momentum
$J=0$, 2, and 4 are plotted in panel~(a) of Fig.~\ref{Mg24:pec}. They are
drawn as a function of the scaled quadrupole moment $\beta_2$ defined as
 \begin{equation}
 \label{beta-q}
 \displaystyle\beta_2
 = \sqrt{\frac{5}{16\pi}}\frac{4\pi}{3R^2A}
   \langle q | 2\hat{z}^2 - \hat{x}^2 - \hat{y}^2 |q \rangle \, ,
 \end{equation}
where $R = 1.2 A^{1/3} \, \text{fm}$ and which varies from $-0.9$ to $+1.6$.
The energies of the first GCM states are also indicated by dots centered
at their mean deformations $\bar\beta_{J\mu}$ defined as
\begin{equation}
\bar\beta_{J\mu}
= \sum_q \beta_2(q) \, |g_{\mu,q}^{J}|^2 \, .
\end{equation}
Although $\bar\beta_{J\mu}$ is not an observable, in axial calculations
it often provides a good indication about the dominant mean-field
configurations in a GCM state.

The corresponding collective wave functions are shown in panel~(b) of
Fig.~\ref{Mg24:pec}.
The $0^+_1$, $2^+_1$ and $4^+_1$ states are a mixing of projected prolate
and oblate deformed configurations, with a dominance of the prolate ones.

\begin{figure}[t!]
\begin{center}
\includegraphics[width=8.4cm]{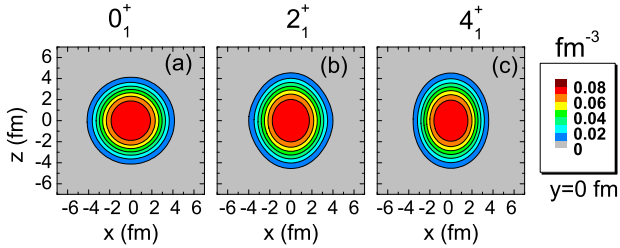}
\end{center}\vspace{-0.5cm}
\caption{\label{Mg24:dens}%
(color online) Contour plots of calculated 3D proton densities
$\rho^{\alpha}_{\alpha} (\br)$ (in fm$^{-3}$) in the $y=0$ plane for the $0^+_1$ (a), $2^+_1$ (b),
$4^+_1$ (c) states (with $M=0$) in $^{24}$Mg.}
\end{figure}

Contour plots of the proton densities $\rho^{\alpha}_{\alpha} (\br)$,
Eq.~\eqref{density:JM}, in the $y=0$ plane are shown in Fig.~\ref{Mg24:dens}
for the $M=0$ orientation of the $J^{\pi}=0^+$, $2^+_1$ and $4^+_1$ states.
As expected, the density of the $0^+_1$ state is spherical after projection,
The densities of the $2^+_1$ and $4^+_1$ states are a superposition of
spherical harmonics with $\lambda$-values ranging from $0$ to $2J$, see Eq.~(\ref{density:JM}). Their elongation along the $z$-axis is larger than along the $x$ and $y$-axes giving to the shapes a prolate-like form. The dimensionless quadrupole deformations $\beta^{(s)}$ determined from the spectroscopic quadrupole moments $Q_s(J_\mu)$ of
$K=0$ states
 \beq
   \label{beta:s}
   \beta^{(s)}(J_k)
   = \sqrt{\displaystyle\frac{5}{16\pi}}\displaystyle\frac{4\pi}{3ZR^2}
     \left(-\dfrac{2J+3}{J} \right) \, Q_s(J_\mu)
 \eeq
are \mbox{$\beta^{(s)}=0.55$} for the $2^+_1$ and $0.63$ for the $4^+_1$
states, respectively. The spectroscopic quadrupole moment $Q_s(J_\mu)$ is
given by the expectation value of the quadrupole operator
$\hat{Q}_{20}(\br) = r^2 \, Y_{20}(\hat{\br})$, multiplied by a
coefficient $\sqrt{16\pi/5}$
\begin{eqnarray}
\label{specQ}
Q_s(J_\mu)
& = & \sqrt{\displaystyle\frac{16\pi}{5}}
      \langle J J 2 0 | J J \rangle \, M^{J \mu}_{J \mu,2}
\, ,
\end{eqnarray}
with $M^{J \mu}_{J \mu,2}$ as defined in Eq.~\eqref{TME}.

\begin{figure}[t!]
\begin{center}
\includegraphics[clip=,width=0.50\textwidth]{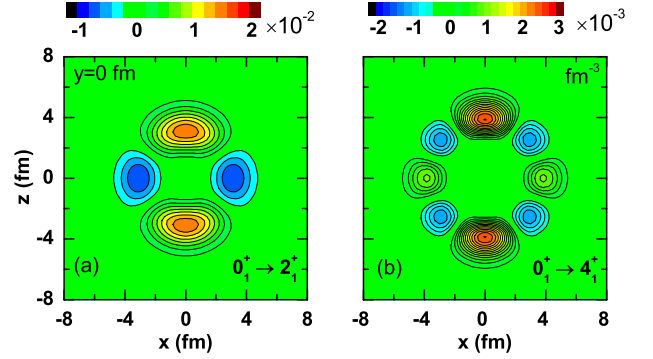}
\end{center}\vspace{-0.5cm}
\caption{\label{Mg24:tdens}%
(color online)
Contour plots of calculated TPD $\rho^{\alpha_f}_{0^+_1}(\br)$,
Eq.~\eqref{gstoes}, in fm$^{-3}$ in the $y=0$ plane for the
inelastic scattering from the ground state to the $2^+_1$ (a) and
the $4^+_1$ (b) states with $M=0$ in $^{24}$Mg.
}
\end{figure}

Figure~\ref{Mg24:tdens} displays the transition proton density (TPD) $\rho^{\alpha_f}_{0^+_1}(\br)$, cf.\ Eq.~\eqref{gstoes},
for the inelastic scattering from the ground state to the $2^+_1$ and $4^+_1$
states of $^{24}$Mg. The density for the transition from the
$0^+_1$ ground state to the $4^+_1$ state is an order of magnitude smaller
than the one to the $2^+_1$ state. As expected from Eq.~\eqref{gstoes}, the
angular part of the TPDs has the shape of a spherical harmonic. They are
the largest around the nuclear surface and present lobes of alternating
signs.

\begin{figure}[t!]
\begin{center}
\includegraphics[width=0.40\textwidth]{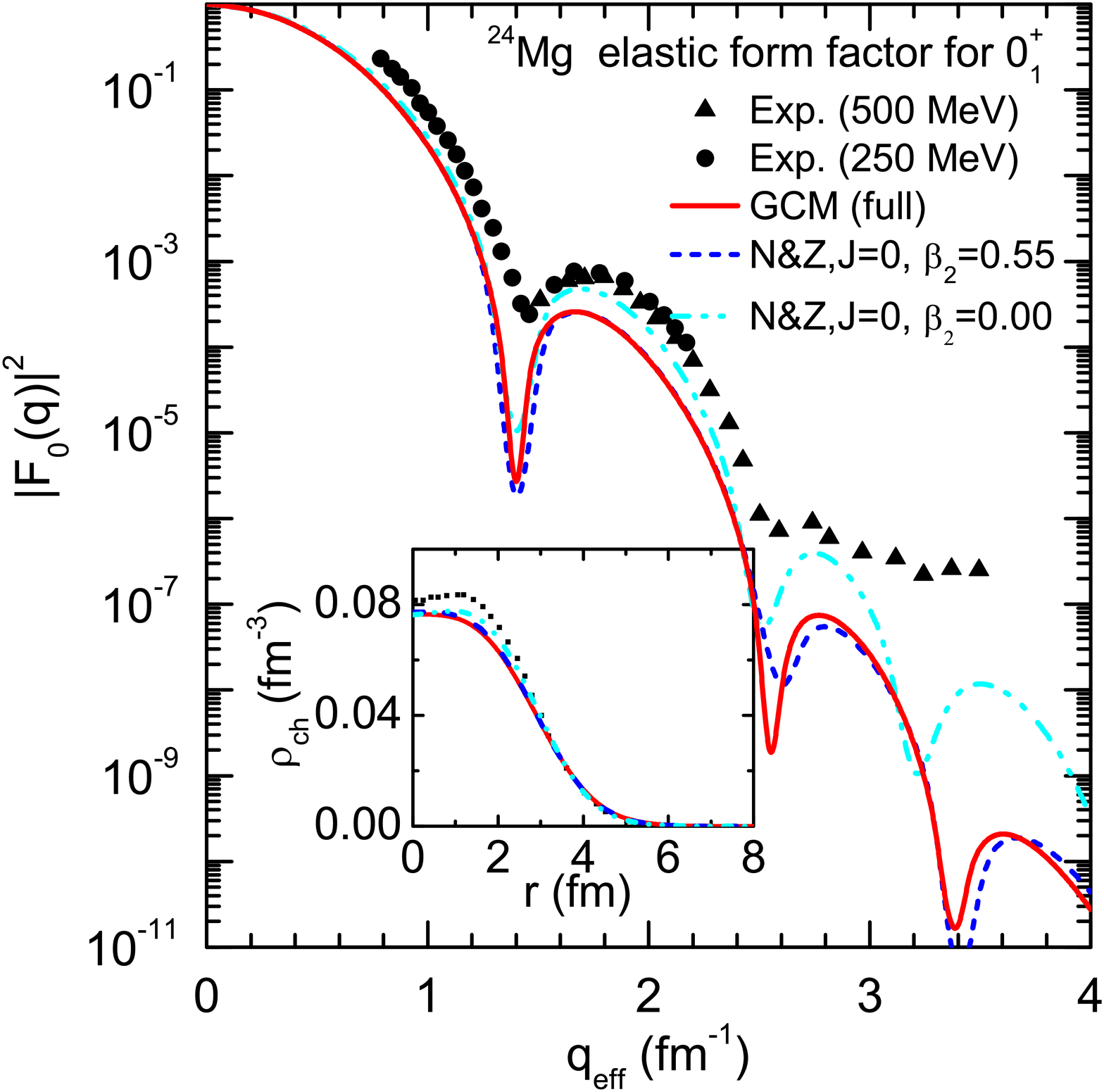}
\end{center}\vspace{-0.5cm}
\caption{\label{Mg24:F0}%
(color online) Elastic C0 form factor $| F_0(q) |^2$ for the $0^+_1$
ground state of $^{24}$Mg, in comparison with several calculations:
The C0 form factor obtained by particle-number and $J=0$ projection of
a single HFB state with either $\beta_2=0$ (spherical shape, light blue
dash-dotted curve) or $\beta_2=0.55$ (minimum of $J=0$ projected energy
curve, dark blue dashed curve) and from the full projected GCM calculation
(red solid curve). The inset shows the corresponding charge density.
Data (solid triangles and circles) taken from Ref.~\cite{Li74}.}
\end{figure}

The elastic C0 form factor $| F_0(q) |^2$ for the ground state of
$^{24}$Mg is plotted in Fig.~\ref{Mg24:F0}. The GCM calculation reproduces the position of
the form factor minima and is in agreement with the data at low q-values. However, our result underestimates
largely the form factor after the first minimum. A similar discrepancy was found in Ref.~\cite{Fuk13} in the case of $^{12}$C.
There, it has been argued that the spreading of the collective wave function on many deformations creates a too large smoothing
of the one-body density and decreases the weights of the large-$q$ components of the transition density. In the case of $^{12}$C, the pure HF form factor was slightly in better agreement with the data. To estimate the effect of deformation
on the form factors, we also show the results obtained from single-configuration calculations
based on either $\beta_2=0$ (spherical shape) or $\beta_2=0.55$ (minimum
of the $J=0$ energy curve) wave functions. The form factor corresponding to the projection of the deformed configuration
differs only marginally from the GCM result. A similar result has also been found for $^{46}$Ar in Ref.~\cite{Wu14a}. On the contrary, the high-$q$ components of the form factor based on the spherical
configuration are much larger and in better agreement with the data.
As can be seen from the inset, the charge density
of the spherical configuration is also larger in the interior than the
densities obtained from $J=0$ projected deformed configurations. Since it is well established that
$^{24}$Mg is deformed, the discrepancy between the GCM result and experiment at large $q$-values points
towards missing components in the ground-state wave function.

\begin{figure}[t!]
\begin{center}
\includegraphics[width=0.40\textwidth]{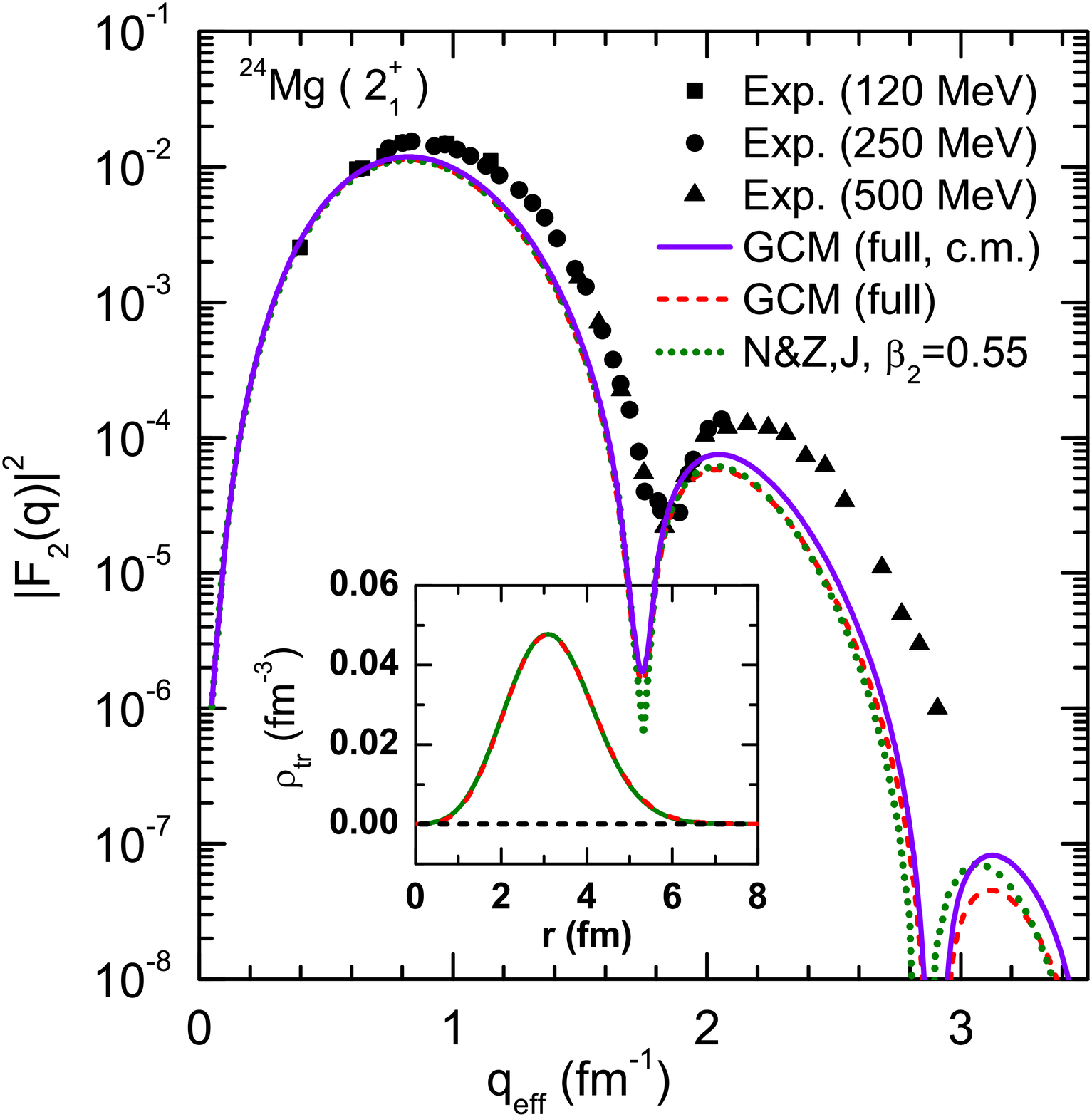}
\end{center}\vspace{-0.5cm}
\caption{\label{Mg24:F2}%
(color online) Longitudinal C2 form factor $| F_2(q) |^2$ for the
transition from the ground state to the $2^+_1$ state for $^{24}$Mg, in
comparison with available data.
The form factor calculated with only
one single configuration of $\beta_2=0.55$ and the form factor of
transition proton density from full GCM calculations are given for
comparison. The inset shows the corresponding transition densities.
Data are taken from Ref.~\cite{Johnston74} (squares) and Ref.~\cite{Li74}
(circles and triangles).}
\end{figure}

In Refs.~\cite{FV82,Friedrich86}, Friedrich and collaborators have
performed a detailed analysis of the relation between various parametric
forms of charge density distributions and the resulting form factors.
They conclude that the first zero of $|F_0(q)|^2$ determines an extension parameter of the charge distribution.
Indeed, their analysis shows that, when comparing two different C0 form
factors, a minimum at lower $q$-values corresponds to a larger extension
of the nuclear density. By contrast, the surface diffuseness of the charge
distribution is related to the height of the first maximum of $| F_0(q)|^2$.
For each of the three calculations shown in Fig.~\ref{Mg24:F0}, the first
minimum of $| F_0(q) |^2$ is located at nearly the same value of $q$,
indicating similar extensions. The value of $| F_0(q) |^2$ at the
first maximum, however, is significantly larger for the spherical
configuration and corresponds to a lower surface thickness, as can be seen
on the plot of the density.

The C2 longitudinal inelastic form factor is plotted in Fig.~\ref{Mg24:F2}
for the transition from the ground state to the $2^+_1$ state in $^{24}$Mg.
Results obtained by projecting a single deformed HFB state with $\beta_2=0.55$
on $J=0$ and $J=2$ are compared with the full projected GCM calculation
and with experimental data. The spreading of the GCM wave function over deformation has little effect.
As for $| F_0(q) |^2$, the GCM $| F_2(q) |^2$ form factor is too low at large $q$-values. A possible cause for this deficiency could be a lack of components not included in the mean-field basis. However, since we are using effective interactions, a shortfall of the EDF cannot be excluded either. To estimate the spurious effect of the COM motion, we have introduced a correction in the form given by Eq.~\eqref{CM}. Although too small, this correction is going into the right direction.

\begin{figure}[t!]
\begin{center}
\includegraphics[width=0.40\textwidth]{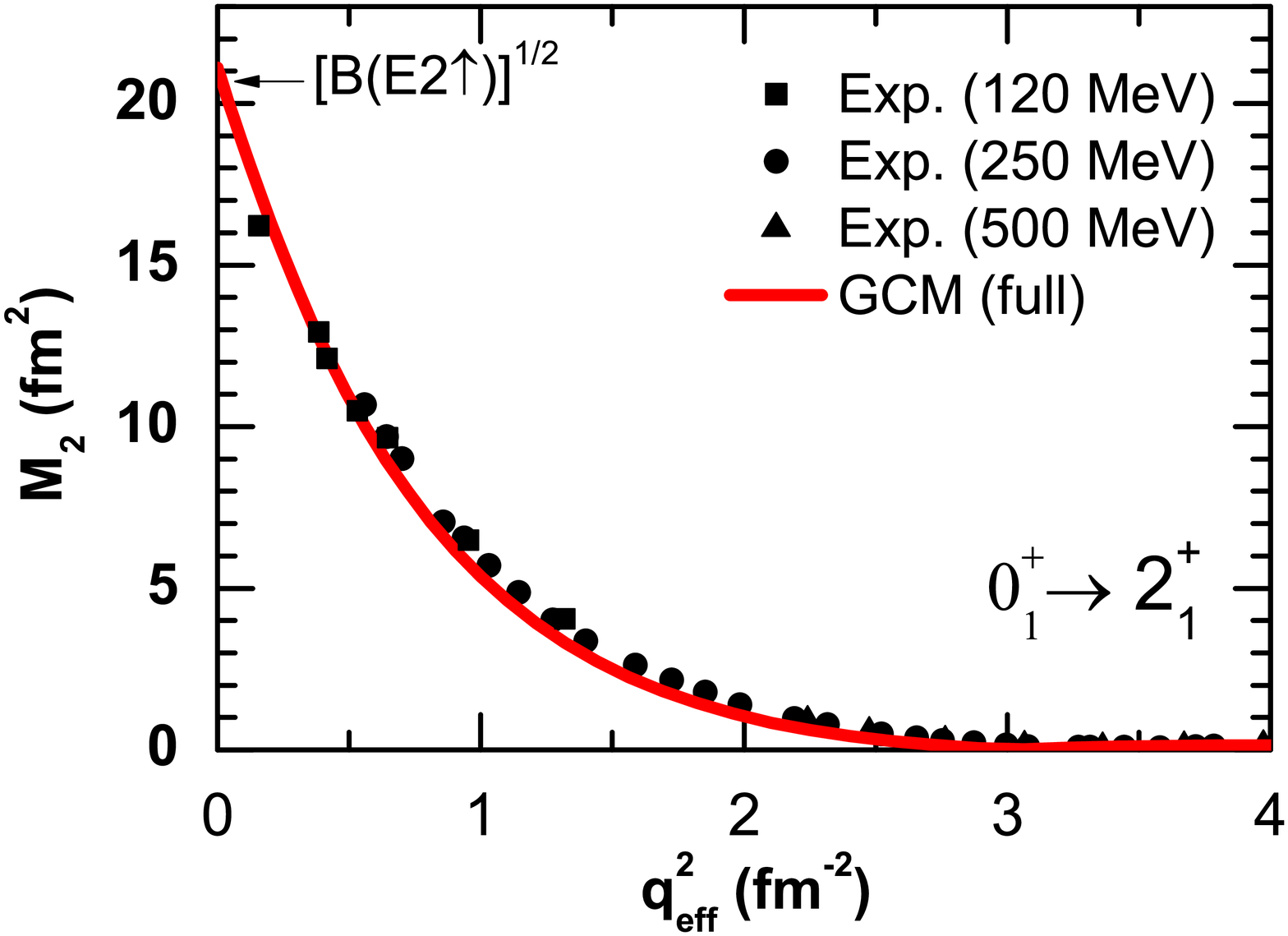}
\end{center}\vspace{-0.5cm}
\caption{\label{Mg24:M2}%
(color online)
$q$-dependent transition quadrupole matrix element $M_2(q^2)$,
Eq.~\eqref{MLq}, for the $E2$ transition from the ground state to the
$2^+_1$ state in $^{24}$Mg, in comparison with available data.
The $M_2(q^2)$ in the $q\to 0$ limit is related to the B(E2) value
via $M_2(0)=\sqrt{B(E2)}/e$.
Data are taken from Ref.~\cite{Johnston74} (squares) and Ref.~\cite{Li74}
(circles and triangles).}
\end{figure}

Figure~\ref{Mg24:M2} displays the $q$-dependent transition quadrupole
matrix element $M_2(q^2)$, Eq.~\eqref{MLq}, for the transition from the
ground state to the $2^+_1$ state. The calculated values
agree well with the available data. According to Eqs.~\eqref{FL}
and~\eqref{MLq}, the transition strength $B(E2)$ is given by the
square of $M_2(q^2)$ in the $q\to 0$ limit. The $B(E2\uparrow)$ value
determined in this way from the inelastic scattering data at low-$q$
region is $420 \pm 25 \, e^2 \, \text{fm}^4$~\cite{Johnston74}, which
is slightly overestimated by our calculation that gives a value of about
$450 \, e^2 \, \text{fm}^4$.

\begin{figure}[t!]
\begin{center}
\includegraphics[width=0.40\textwidth]{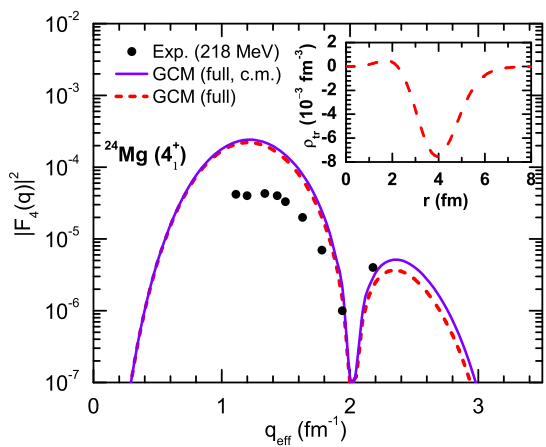}
\end{center}\vspace{-0.5cm}
\caption{\label{Mg24:F4}
(color online) Longitudinal C4 form factor $| F_4(q) |^2$ for the
transition from the ground state to the $4^+_1$ state of $^{24}$Mg, in
comparison with available data. The inset shows the corresponding transition density.
Data are taken from Ref.~\cite{Zarek78}.}
\end{figure}

Figure~\ref{Mg24:F4} displays the C4 longitudinal inelastic form factor $| F_4(q) |^2$ from the ground state to the $4^+_1$ state. The experimental data are taken from Ref.~\cite{Zarek78}. The calculation reproduces well the diffraction minimum observed at $q \simeq 2.0 \, \text{fm}^{-1}$ in the data. Moreover, the calculated $E4$ transition strength
$B(E4: 0^+_1 \to 4^+_1)=2.07 \times 10^3 \, e^2 \, \text{fm}^8$ is close to the experimental value of $2.0(3) \times 10^3 \, e^2 \, \text{fm}^8$~\cite{Zarek78}. The $L=4$ transition density, shown in the inset of Fig.~\ref{Mg24:F4}, is peaked at $r\simeq4.0$ fm, further out than the $L=2$ transition density that has been shown in Fig.~\ref{Mg24:F2}.


\section{Application to even-mass $^{58-68}$N\lowercase{i}}
\label{Sec.IV}

The stable Ni isotopes ($A=58$ to 62) have been extensively studied in the 1960's. The data have been extended to heavier
isotopes over the last ten years, going up to potentially neutron magic
numbers $N=40$ and $N=50$. There is now a large set of data putting into evidence the complexity of the evolution of the Ni shell structure with the number of neutrons (see for instance the discussions in Refs.~\cite{Orce08,Broda12,Allmond14}).

\begin{figure}[t!]
\begin{center}
\includegraphics[width=8.5cm]{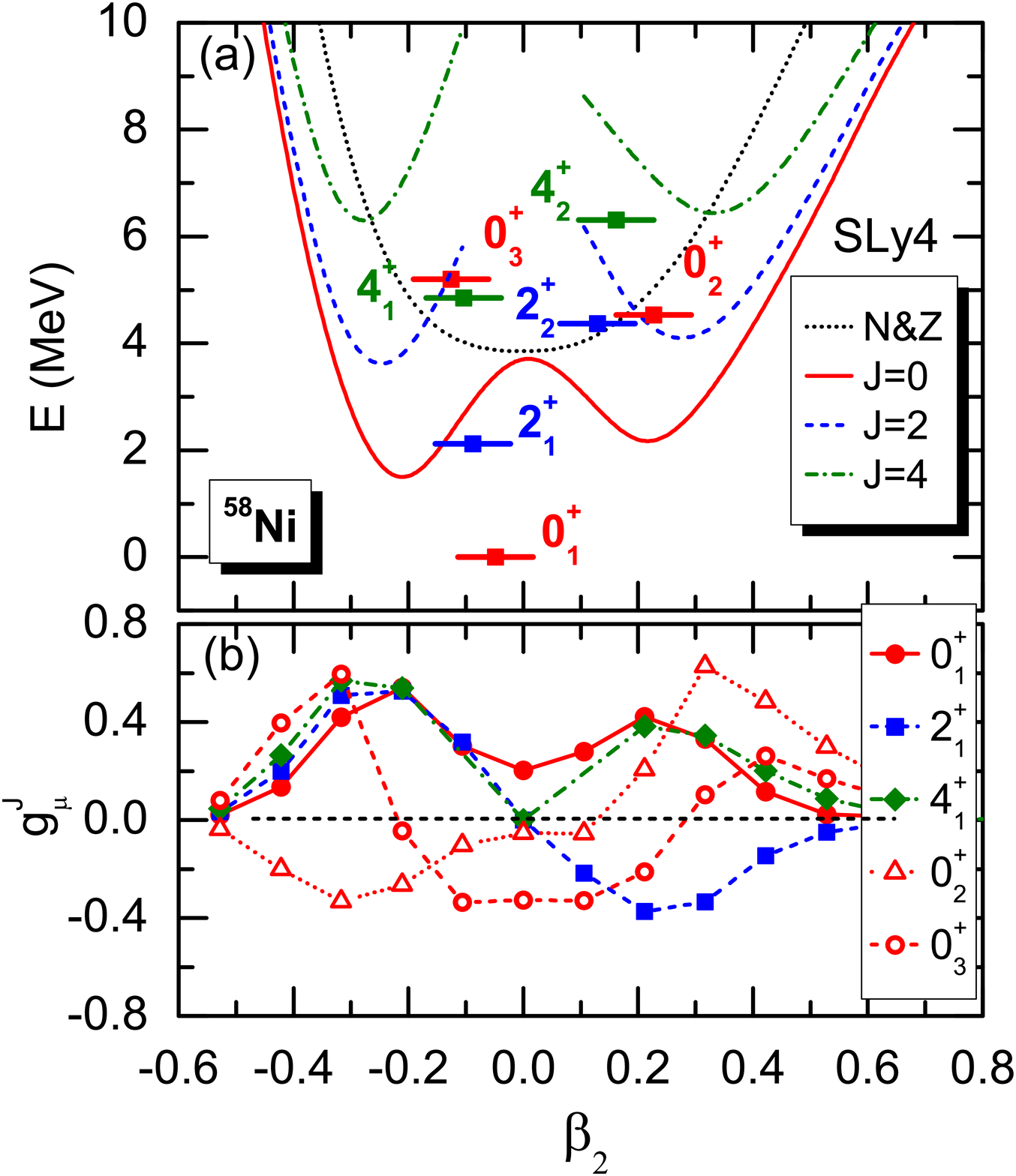}
\end{center}\vspace{-0.5cm}
\caption{\label{Ni58:spec}%
(color online) Same as Fig.~\ref{Mg24:pec}, but for $^{58}$Ni.}
\end{figure}

\begin{figure}[t!]
\begin{center}
\includegraphics[width=8.5cm]{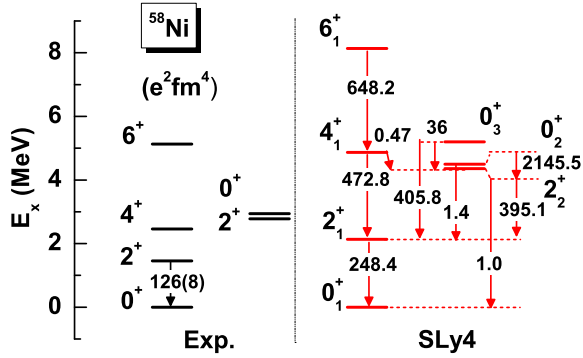}
\end{center}\vspace{-0.5cm}
\caption{\label{Ni58:spec2}%
(color online) Comparison between the spectrum obtained for $^{58}$Ni using our method and the experimental results. Data are taken from Ref.~\cite{Allmond14}.}
\end{figure}

\begin{figure}[t!]
\begin{center}
\includegraphics[width=8.5cm]{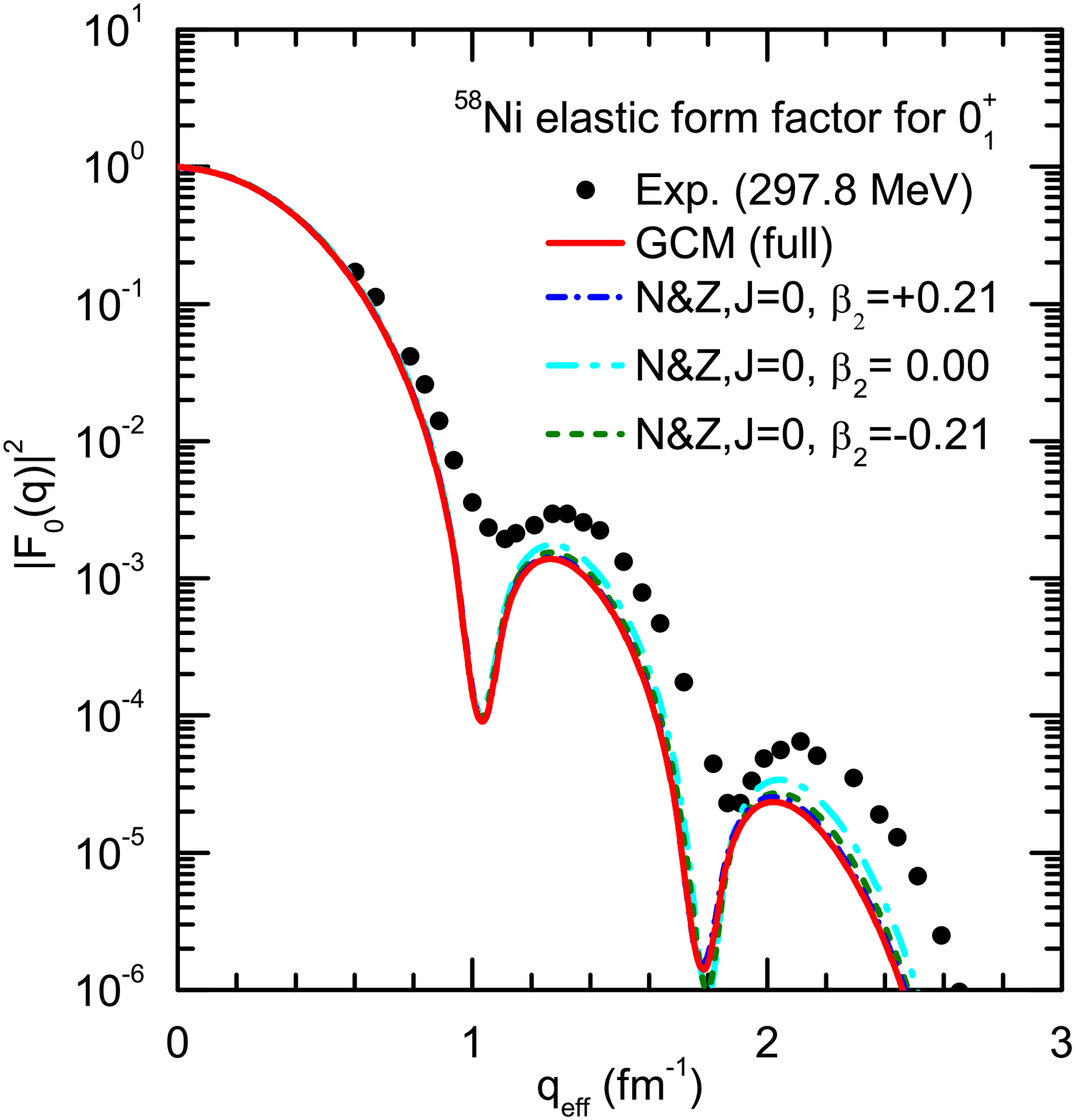}
\end{center}\vspace{-0.5cm}
\caption{\label{Ni58:F0}%
(color online) Data for the elastic form factor $| F_0(q) |^2$ for the
ground state of $^{58}$Ni taken from Ref.~\cite{Ficenec70} in comparison
with the form factor obtained from four different calculations:
projection of a single HFB configuration with either $\beta_2=0$
(spherical shape) or $\beta_2 = \mp 0.21$ (oblate and prolate minima
of the $J=0$ energy curve) and full GCM of projected states.
}
\end{figure}

The discretization of the mass quadrupole moment used for the GCM calculations of the Ni isotopes ranges from $-500 \; \text{fm}^2$ to $+700 \; \text{fm}^2$ with a step size $\Delta q = 100 \; \text{fm}^2$.

The results obtained at the successive steps of our method for $\elemA{58}{Ni}$ are plotted in Fig.~\ref{Ni58:spec}. The energy curve obtained from particle-number
projection of mean-field wave functions presents a soft spherical minimum.
After projection on angular momentum, two minima, close in
energy, are obtained for $J=0$, $2$ and $4$ by the projection of prolate and oblate mean-field configurations.
The collective wave functions $g^J_\mu$ resulting from configuration mixing are spread over a large range of deformations
(see panel (b) of Fig.~\ref{Ni58:spec}).  As a consequence, the mean deformation $\bar{\beta}_{J\mu}$  is close to zero and does not bring valuable information. By contrast, the positive $\bar{\beta}_{J\mu}$ of the first excited states indicates that they are dominated by prolate mean-field configurations.

There have been claims in the literature that $\elemA{58}{Ni}$
is a spherical vibrator, see for instance Refs.~\cite{Stefanini95,Sim02}.
The calculated energy pattern that we obtain, shown in Fig.~\ref{Ni58:spec2},
has indeed some of the characteristics expected for a
vibrator~\cite{Garrett10a}. The first $4^+$ and second $2^+$ levels are
at about two times the energy of the $2^+_1$ state. There are, however,
two near-degenerate $0^+$ levels at the expected energy of the two-phonon
state instead of just one.
Looking at transition probabilities, the first excited $0^+$ has a strong
deexcitation to the second $2^+$ state, which is incompatible with
a simple vibrator. The second excited $0^+$ decays predominantly to
the first $2^+$ and would thus be a better candidate for a two-phonon state,
but the overall pattern of $B(E2)$ values is very different from what would
be expected. The available data for $\elemA{58}{Ni}$ are too sparse to
draw firm conclusions, but they do not seem to be well described by the
assumption of a simple vibrator either. In fact, there seems to be a
general rule that the more information about a potential
anharmonic vibrator becomes available, the less this interpretation can be
retained \cite{Garrett08a,Garrett10a}.

\begin{figure}[t!]
\begin{center}
\includegraphics[width=0.45\textwidth]{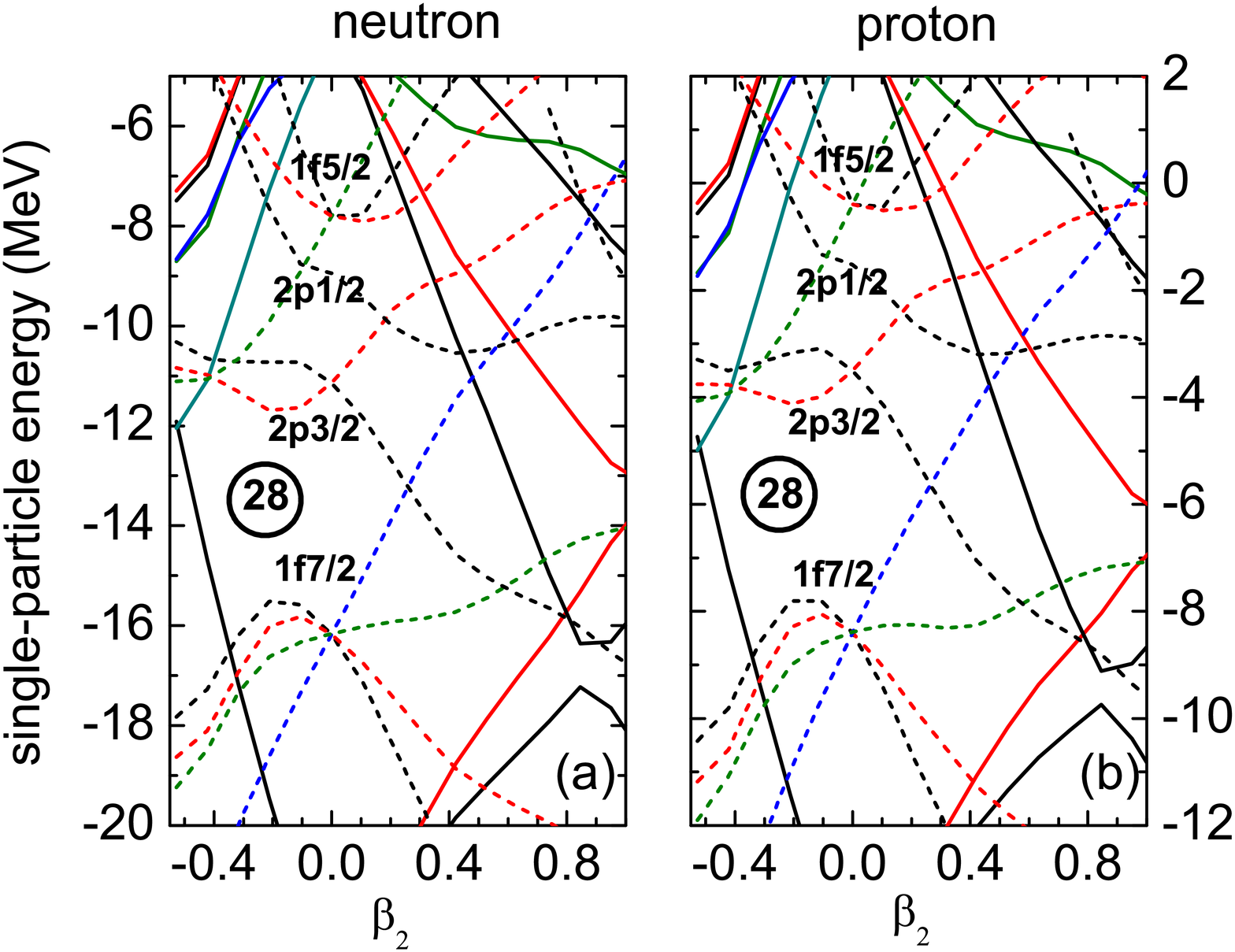}
\end{center}
\vskip-0.6cm
\caption{\label{spe}%
(color online)
Nilsson diagram of the eigenvalues of the single-particle Hamiltonian
for neutrons (a) and protons (b) as obtained with the Skyrme interaction SLy4 as a function of the quadrupole deformation. Solid (dotted) lines represent levels of positive
(negative) parity, and black, red, green and blue color represents
levels with expectation values of $\langle j_z \rangle = 1/2$, $3/2$, $5/2$
and $7/2$.}
\end{figure}

The shell-model description of this Ni isotope, and also of all
others up to $\elemA{68}{Ni}$, shows that a correct reproduction
of both energies and B(E2) values of the low-lying states requires
to include the full $fp$-shell, see the discussion in Ref.~\cite{Broda12}. In its present form,
our beyond mean-field method does not allow to include all the
relevant shell-model configurations: multiple quasiparticle
excitations that break time reversal invariance are not contained in the
model space used in this study. However, deformed configurations include
many spherical multi-particle-multi-hole excitations. The spreading of the GCM
wave functions over a large range of deformations is an economic
way to include spherical orbitals arising from shells excited
at sphericity (see Fig.~\ref{spe}).

The elastic scattering form factor for $^{58}$Ni is shown in
Fig.~\ref{Ni58:F0}. The results obtained with the full GCM basis are compared to those corresponding to the projection of a single configuration,
either spherical or corresponding to the mimima at $\beta_2=\pm 0.21$ of the
projected energy curve. All these form factors are quite close, with slight differences
at $q$-values beyond the first maximum. The position of the zeros is reproduced rather
well, but the heights of the first and second
diffraction maxima are underestimated.

\begin{figure}[t]
\begin{center}
\includegraphics[clip=,width=0.40\textwidth]{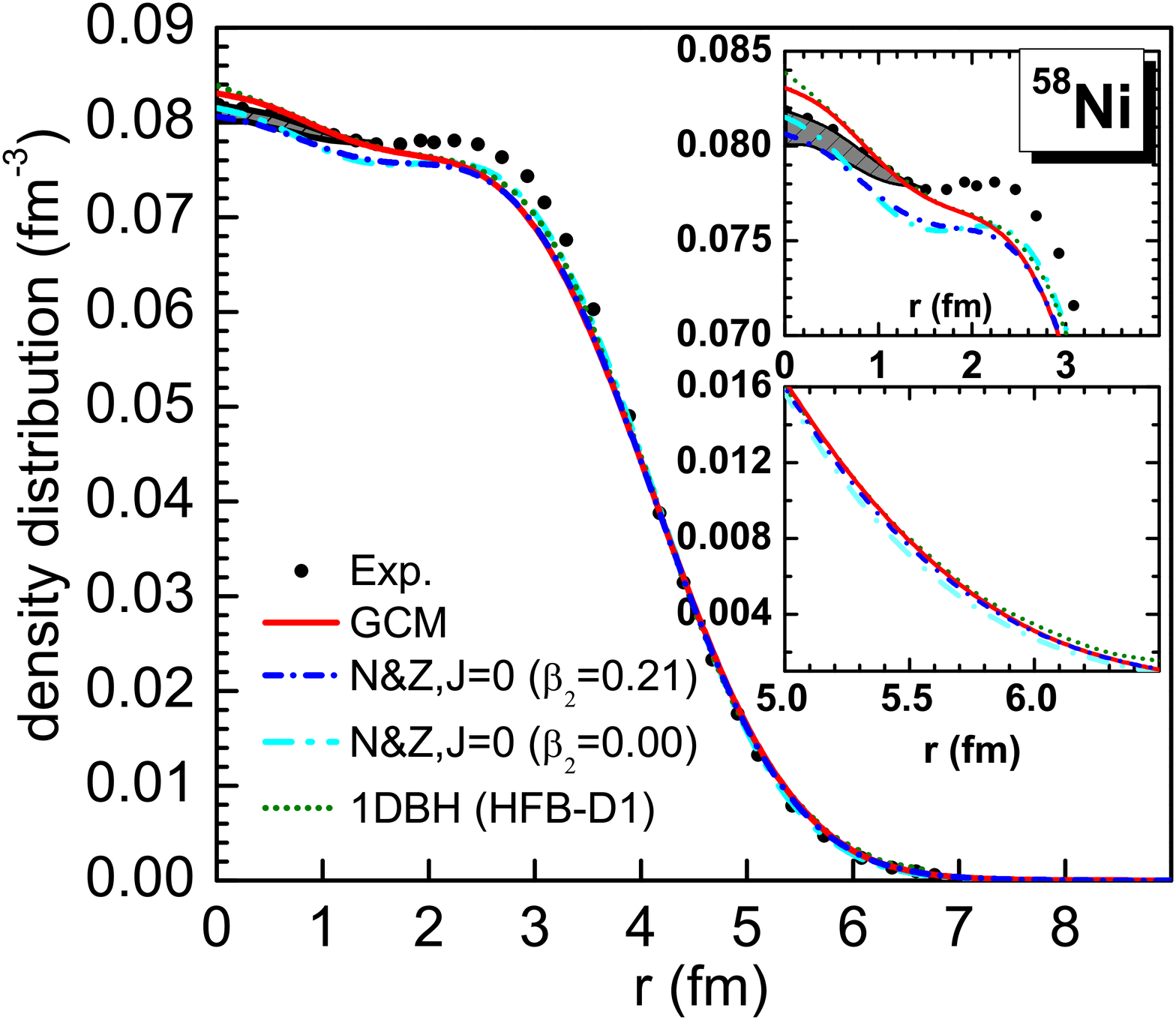}
\end{center}\vspace{-0.5cm}
 \caption{\label{Ni58:dens}%
(color online) Comparison between the charge distribution of the
ground state of $^{58}$Ni obtained using single projected mean-field
configurations or the full GCM basis and the experimental data~\cite{Sick75}.
A previous calculation using a one-dimensional Bohr Hamiltonian based on
an HFB calculation with the Gogny D1 force (1DBH)~\cite{Girod76}
is also shown. The insets magnify the profile of the charge density
at very small radii and in the nuclear surface.}
\end{figure}

The ground-state charge density distribution
is plotted in Fig.~\ref{Ni58:dens} for the same four calculations
as in Fig.~\ref{Ni58:F0}. The small differences between these
calculations above $q=1.2 \, \text{fm}^{-1}$ is reflected in differences
between the densities in the interior region ($r<2.0 \, \text{fm}$).
The GCM result is similar to a previous result obtained from a
one-dimensional Bohr Hamiltonian (1DBH) calculation determined by
the HFB method and using the Gogny D1 force~\cite{Girod76}.

\begin{figure}[t!]
\begin{center}
\includegraphics[width=8.5cm]{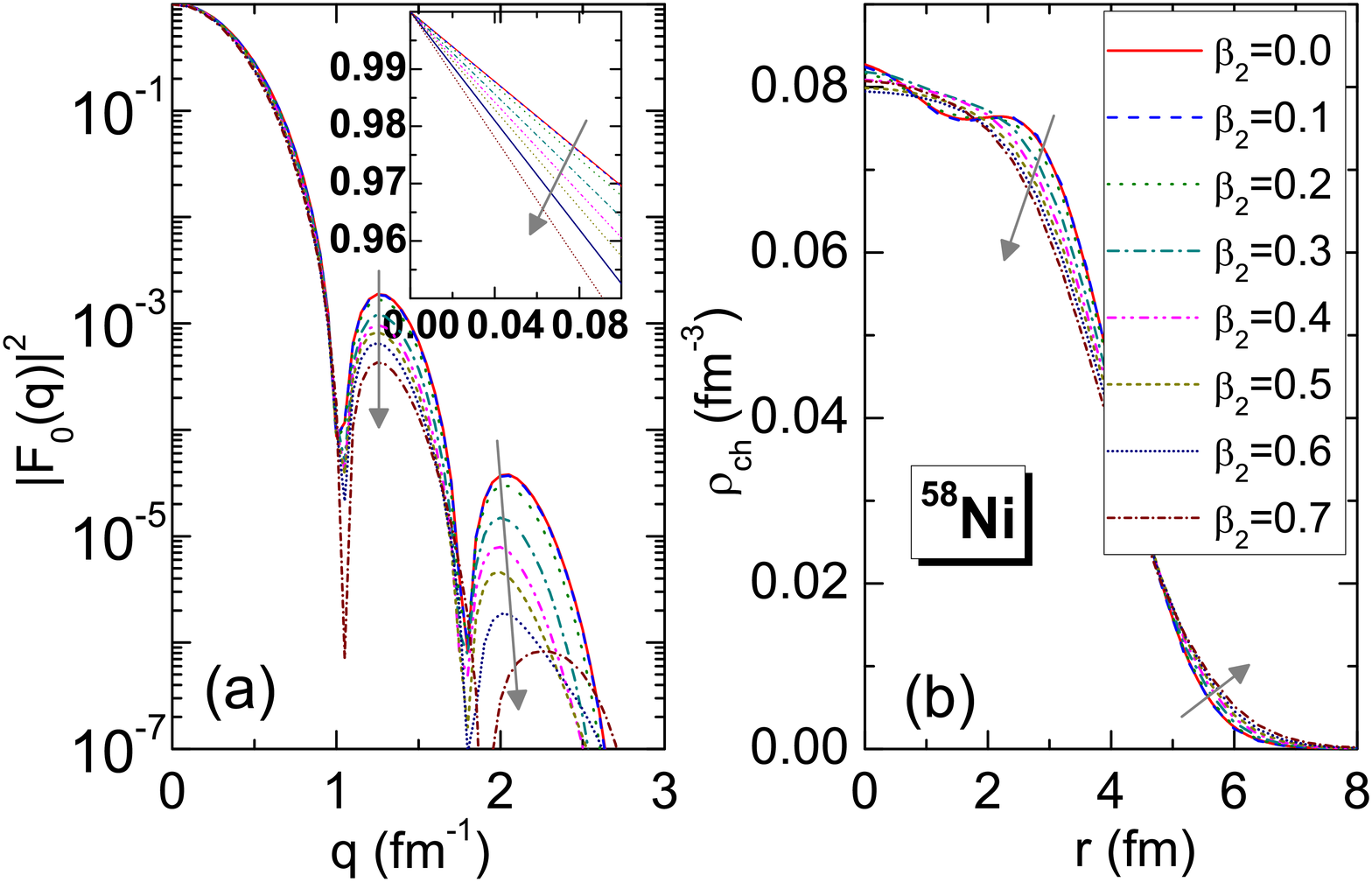}
\end{center}\vspace{-0.5cm}
\caption{\label{Ni58:deformation}%
(color online) (a) Elastic form factor $| F_0(q)|^2$ for the
$J=0$ state of $^{58}$Ni projected from a single HFB configuration with prolate
deformation of  $\beta_2$ increasing from $0.0$ to $0.7$, respectively.
(b) Charge distributions corresponding to the form factors
displayed in panel (a).}
\end{figure}

To analyze the effect of static deformations, we compare in
Fig.~\ref{Ni58:deformation} the elastic form factor and the
charge distribution  calculated using projected deformed configurations with increasing values of $\beta_2$
from spherical to $\beta_2=0.7$. The height
of the first and second diffraction maxima  is not affected by
small deformations. However, it starts to significantly decrease with deformation for $\beta_2$ values larger than
$0.2$. Moreover, the C0 form factor drops faster
in the low-$q$ region if the deformation is increased, as shown in the inset of Fig.~\ref{Ni58:deformation}.
This behaviour can be understood by looking to the relation~(\ref{F0}) between the C0 form factor and the rms charge radii $r_{\text{ch}}$ for low-$q$ values and from the effect of deformation on the charge radius of a
uniformly charged liquid drop,
$r_{\text{ch}}/r^{\text{sph}}_{\text{ch}} \simeq \big( 1+\frac{5}{4\pi} \, \beta^2_2 \big)$.
Panel (b) of Fig.~\ref{Ni58:deformation} illustrates the effect of deformation on the charge density
distribution. Increasing the deformation pushes charge from the inside of the
surface (around $r=3 \, \text{fm}$) to the outside
(around $r=6 \, \text{fm}$).

\begin{figure}[t!]
\begin{center}
\includegraphics[width=0.40\textwidth]{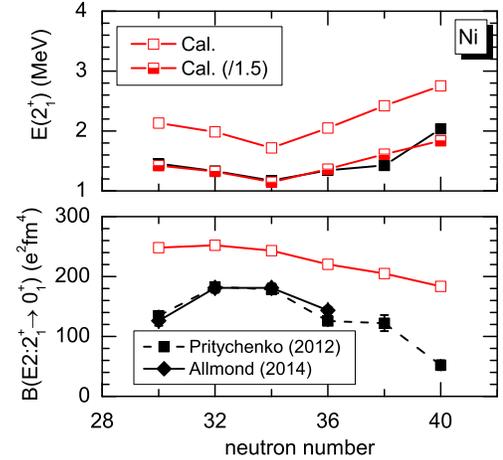}
\end{center}\vspace{-0.5cm}
\caption{\label{Ni:BE2}%
(color online)  The excitation energy $E(2^+_1)$ of $2^+_1$ state and
the $B(E2)$ value for the transition from this state to the ground state
for $^{58-68}$Ni.
Experimental data are taken from Refs.~\cite{Pritychenko11,Allmond14}.
}
\end{figure}

\begin{figure}[t!]
\begin{center}
\includegraphics[width=0.50\textwidth]{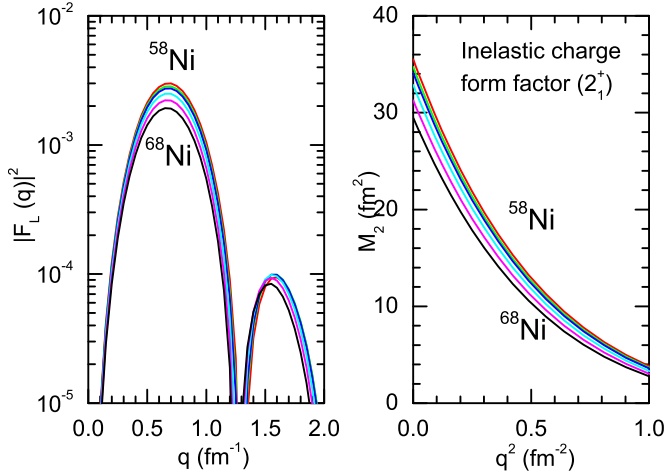}
\end{center}\vspace{-0.5cm}
\caption{\label{Ni:C2}
(color online)
(a) The C2 form factor and (b) $q$-dependent transition
 quadrupole matrix element for the quadrupole transition from ground state
to $2^+_1$ state in $^{58-68}$Ni.}
\end{figure}

\begin{figure}[t!]
\begin{center}
\includegraphics[width=0.50\textwidth]{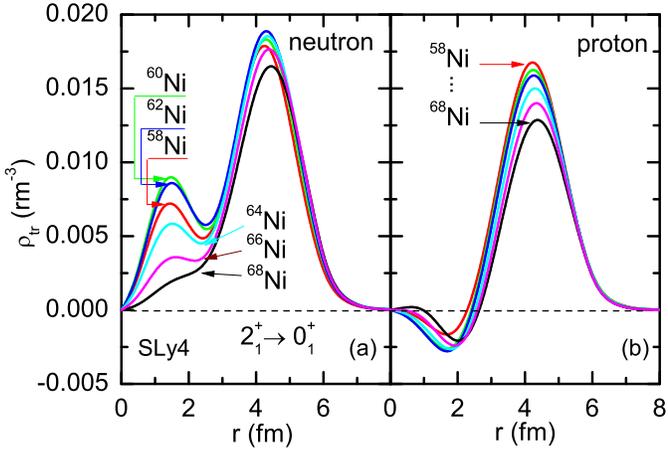}
\end{center}\vspace{-0.5cm}
\caption{\label{Ni:tdens2}%
(color online)
Calculated transition neutron and proton densities from the $2^+_1$ state
to the ground state for $^{58-68}$Ni. }
\end{figure}

The origin of the change of behaviour of $| F_0(q) |^2$ at $\beta_2 \approx 0.3$ can be traced back to
 the single-particle spectra. These are plotted in Fig.~\ref{spe}. The shell structure for neutrons and for protons
is very similar. At $\beta_2 \approx 0.3$, a downsloping proton level from the $1f_{7/2}$ spherical shell
crosses an upsloping level from the $2p_{3/2}$ shell. It indicates that the gradual population of the $2p_{3/2}$ orbital beyond this point might be
responsible for the the decrease of the form factor at large $q$-values.

In the next figures, we show results obtained for the even Ni isotopes up to $N=40$. Figure~\ref{Ni:BE2} shows the evolution with $N$ of the excitation energy  of the first $2^+_1$ state and of the $B(E2)$ value to the ground state. Although both the $E(2^+_1)$ and $B(E2)$ values are systematically overestimated in our calculation, their evolution as a function of the neutron number is rather well reproduced. We expect that the discrepancy with experiment is mainly due to the time-reversal invariance that is imposed to the mean-field wave functions and that limits the model space of the present calculation to purely collective states. Non-collective time-reversal-invariance-breaking 2-qp excitations are indeed present in the shell model calculations that are in better agreement with data. It can be expected that such configurations will decrease the $2^+$ excitation energies and make them less collective, resulting in a decrease of the B(E2)-values.

 The calculated C2 form factor $|F_2(q)|^2$ and the $q$-dependent transition quadrupole matrix element $M_2(q^2)$ for the quadrupole transition from the ground state to the $2^+_1$ state are displayed in Fig.~\ref{Ni:C2}. The isotopic dependence of the form factor is very weak, with a decrease of the height of the first maximum with $N$. The quadrupole transition matrix element $M_2(q)$ at $q\to0$ decreases in the same way, which corresponds to the smooth decrease of the calculated $B(E2)$ value, cf.\ Fig.~\ref{Ni:BE2}.

\begin{figure}[]
\begin{center}
\includegraphics[clip=,width=0.45\textwidth]{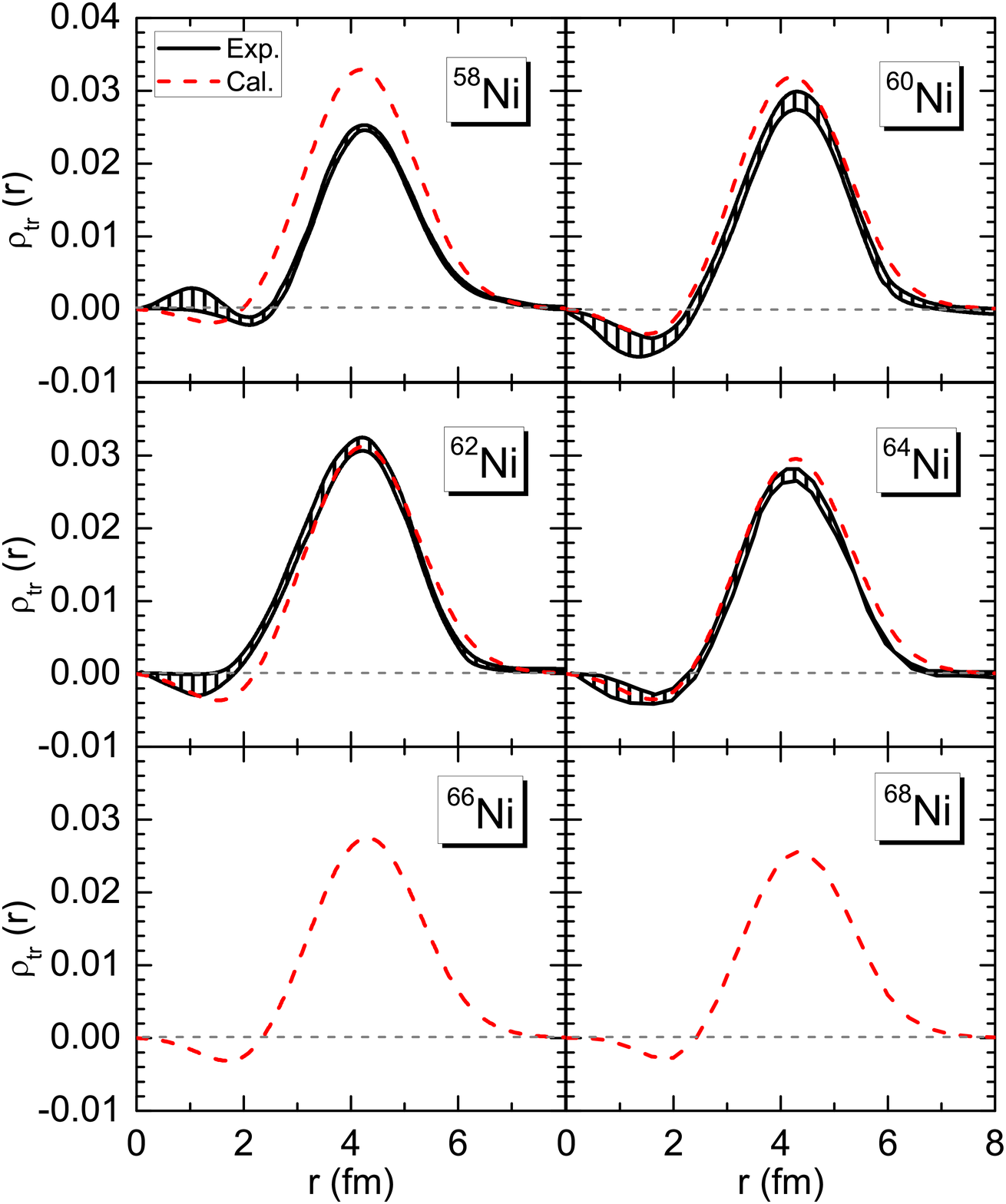}
\end{center}\vspace{-0.5cm}
\caption{\label{Ni:tdens3}%
color online) Calculated transition charge densities from the ground state
to $2^+_1$ state for $^{58-68}$Ni, in comparison with available
data~\cite{Heisenberg81}. }
\end{figure}

In Fig.~\ref{Ni:tdens2}, the neutron and proton densities for the transition from the $2^+_1$ state to the ground state are shown for $^{58-64}$Ni. The radial profiles are similar for all isotopes with a large peak at large radii and a smaller one at low values of $r$. The height of the first peak for the neutron transition density decreases with $N$, and nearly disappears at $N=40$, in contrast with the second peak.

 The ratio $\eta$ between the values of the quadrupole matrix element for neutrons to that for protons is given in Tab.~\ref{tab1} for $^{58-68}$Ni.
 This ratio provides a measure of the isovector character of the  transition. It is close to one in our calculation, indicating that the transitions are predominantly isoscalar.

\begin{table}
\caption{\label{tab1}
%
Isovector character $\eta$ [cf.\ Eq.~\eqref{eta}] of the $2^+_1$ state
of even-even Ni isotopes.
}
\begin{center}
\begin{tabular}{ccccccc}
\hline \hline \noalign{\smallskip}
  $\eta$                & $^{58}$Ni & $^{60}$Ni & $^{62}$Ni & $^{64}$Ni & $^{66}$Ni & $^{68}$Ni\\
\noalign{\smallskip}  \hline \noalign{\smallskip}
  This work             & 1.02 & 1.05 &  1.06 & 1.06 &	1.03 &	1.02   \\
  Ref.~\cite{Terrien73} & 1.01 & 1.02 &  1.12 & 0.92 &	  &    \\
  Ref.~\cite{Chaumeaux78} & 1.10 & 1.31 &  1.36 & 1.41 &	  &    \\
  Ref.~\cite{Lombard81} & 1.10 & 1.09 &  1.33 & 1.02 &	  &    \\
\noalign{\smallskip} \hline \hline
\end{tabular}
\end{center}
\end{table}

\begin{figure}[]
\begin{center}
\includegraphics[width=8cm]{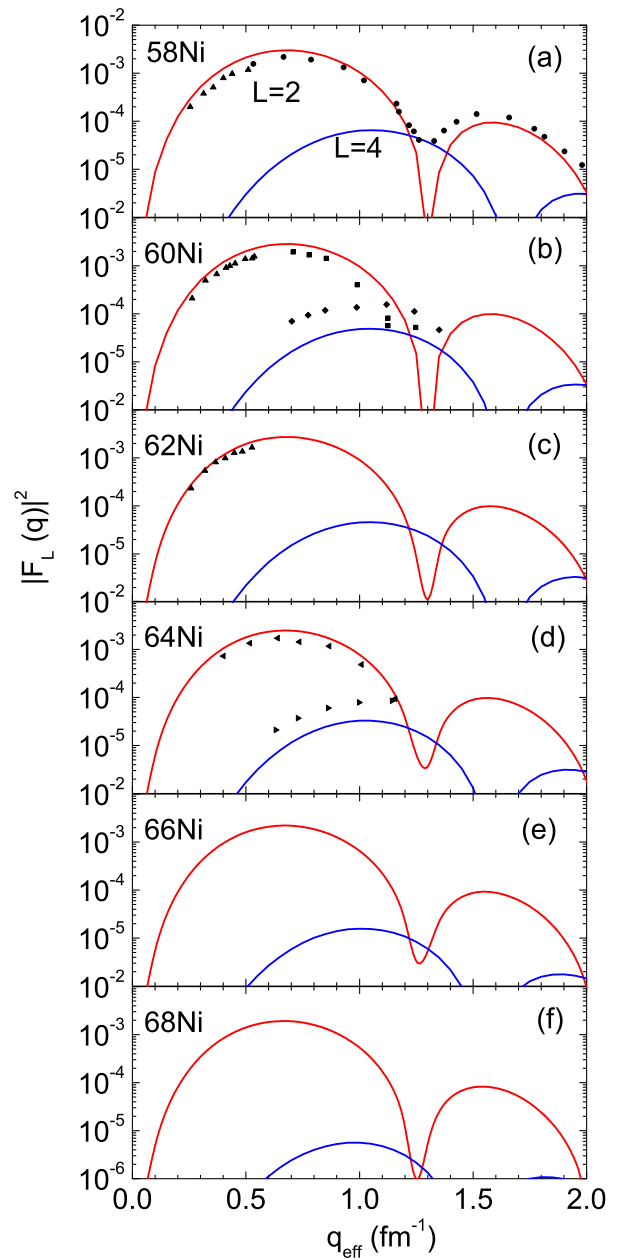}
\end{center}\vspace{-0.5cm}
 \caption{\label{Ni:F24}
(color online) Calculated inelastic Coulomb form factors $| F_L(q)|^2$ for the transition from the ground state to $J^+_1$ $(L=J=2, 4)$ state in $^{58-68}$Ni, in comparison with available data, taken from Ref.~\cite{Duguay67} (up triangles), Ref.~\cite{Torizuka69} (squares and diamonds), Ref.~\cite{Frois83} (circles) and Ref.~\cite{Braunstein88} (left and right triangles). }
\end{figure}

The radial transition charge density (TCD) from the ground state to the $2^+_1$ state is compared to the experimental data~\cite{Heisenberg81} for $^{58-68}$Ni in Fig.~\ref{Ni:tdens3}. The shape of TCD of $^{58-64}$Ni is reproduced by the GCM calculation. However, we overestimate the height of the surface peak and/or the tail part of the TCD. This deficiency can be traced back to the overestimated $B(E2:2^+_1\to 0^+_1)$ values, as shown in Fig.~\ref{Ni:BE2}. In Fig.~\ref{Ni:F24} the GCM inelastic Coulomb form factors $| F_L(q)|^2$ is compared to the experimental data for the transitions from the ground state to $J^+_1$ $(L=J=2, 4)$ state. Our calculation reproduces rather well the shapes of the quadrupole and hexadecapole transition form factors, but systematically underestimates the hexadecapole ones.

\section{Summary and outlook}
\label{Sec.V}

We have presented how to determine densities and transition densities, as well as the corresponding form factors, within the beyond mean-field model that we develop since many years. The light deformed nucleus $^{24}$Mg and the even-mass $^{58-68}$Ni have been used as examples. Depending on the structure of the nucleus, static deformation, or dynamic
shape fluctuations, or both, might be important for the description of the ground-state and transition densities.

The framework that we have developed is very general and can be applied to any nucleus and any kind of transitions for which calculations using the GCM are available. This gives some hope that applications to odd-mass nuclei will be available in a not too distant future~\cite{Bally14}. For a better description of low-lying excited states in spherical even-even nuclei, it would be desirable to add non-collective time-reversal-breaking $n$-quasiparticle states to the GCM basis.

Leptonic probes have the advantage that the interaction mechanism
and the nucleonic form factors are precisely known, which reduces the
theoretical uncertainties. But with additional modeling, also the
scattering of hadronic probes off nuclei could be described.

To improve the quality of the results obtained in our model, one certainly needs to construct a
new energy functional, which should be adjusted to the data on nuclear
charge radii at the beyond-mean-field level.
As has been shown in Ref.~\cite{BBH08}, the charge radii, in particular
of light nuclei, become systematically larger in angular-momentum
projected GCM, which poses a problem when using a parametrization adjusted
at the mean-field level. Elastic and inelastic form factors seem to be tools very sensitive
to the momentum composition of the collective wave functions should provide stringent tests of nuclear models.


\begin{acknowledgments}

Fruitful discussions with S.~Baroni, P.-G.~Reinhard and K.~Washiyama
are gratefully acknowledged. We also thank H. Mei for critical checking of the formulae for
transition densities. This research was supported in parts by the PAI-P6-23 of the Belgian Office for Scientific Policy, the F.R.S.-FNRS (Belgium),
the National Science Foundation of China under Grants No. 11305134 and 11105111, by the European Union's Seventh Framework Programme ENSAR under grant agreement n262010, and by the CNRS/IN2P3 through the PICS No.~5994.

\end{acknowledgments}


\begin{appendix}

\section{Form factors of electron scattering off nuclei in PWBA}
\label{append1}

We suppose that the nucleus makes a transition from the initial state
$| \alpha_i \rangle$ to the final state $| \alpha_f\rangle$, where we introduce the shorthand notation $\alpha$ representing $JM\mu$. In the plane-wave Born approximation (PWBA), the longitudinal form factor, normalized to the nuclear charge
$Z$, is given by the Fourier-Bessel transformation of the transition density $\rho^{\alpha_f}_{\alpha_i}(\br)$,
cf.\ Eq.~(\ref{tdens:gcm}),
\beqn
| F (\bq) |^2
& = & \dfrac{4\pi}{Z^2 \, \hat J^{2}_i}
      \sum_{M_i,M_f} \left|  \sum_{LM}
      \langle \alpha_f| \hat{M}_{LM} | \alpha_i \rangle
       Y^\ast_{LM}(\hat \bq) \right|^2 \, .
\eeqn
The multipole operator $\hat{M}_{LM}(q)$ has been defined following
Refs.~\cite{Duguay67,Raphael70},
\beq
\label{multipole-opeator}
\hat M_{LM}(q)
\equiv  \int \! d^3r \; j_L(qr) \, Y_{LM}(\hat\br) \, \hat\rho(\br) \, ,
\eeq
where $j_L(qr)$ is a spherical Bessel function and where
we have used the relation
\beq
e^{\text{i}\bq\cdot\br}
= 4\pi\sum_{LM} \text{i}^L \, j_L(qr) \, Y^\ast_{LM}(\hat \bq) \, Y_{LM}(\hat\br).
\eeq
By using the orthogonality of the spherical harmonics, one can show that the radial dependence of $| F(\bq) |^2$ is given by~\cite{Forest66,Duguay67,Friar85}
\beqn
| F (q) |^2
& = & \dfrac{4\pi}{ Z^2\hat J^2_i} \sum^{\infty}_{L=0}
      \left| \langle J_f \mu_f || \hat{M}_{L}(q) || J_i \mu_i \rangle
      \right|^2 \, .
\eeqn
Comparing with Eq.~\eqref{cross-section2}, one finds the form factor
$F_L(q)$ for an angular momentum transfer $L$~\cite{Raphael70},
\beqn
\label{FF}
F_L(q)
= \dfrac{\sqrt{4\pi}}{Z\hat J_i}
 \left| \langle J_f\mu_f || \hat M_{L}(q) || J_i\mu_i \rangle \right| \, .
\eeqn
Using its definition provided by Eq.~\eqref{multipole-opeator}, the matrix element of the multipole operator $\hat{M}_{LM}$
between an initial $| \alpha_i \rangle$
and a final $| \alpha_f\rangle$ state
\beq
\langle \alpha_f | \hat M_{LM} |\alpha_i \rangle
= \int \! d^3r \; j_L(qr) \, Y_{LM}(\hat\br) \,
      \rho^{\alpha_f}_{\alpha_i}(\br)
\eeq
is related to the reduced matrix element
$\langle J_f; \mu_f || \hat M_{L}(q) || J_i ; \mu_i \rangle$
by the Wigner-Eckart theorem~\cite{Varshalovich88}
\beqn
\label{eq:WET}
\langle J_f \mu_f || \hat M_{L}(q) || J_i \mu_i \rangle
& = & (-1)^{2L}\hat{J_f} \,
      \dfrac{\langle \alpha_f | \hat M_{LM} | \alpha_i\rangle}
            {\langle J_i M_i L M | J_f M_f \rangle} \, ,
      \nonumber \\
\eeqn
where $\langle J_i M_i L M | J_f M_f \rangle$ is a Clebsch-Gordan coefficient.
In other words, one can define a \textit{reduced transition density} $\rho^{J_f,\mu_f}_{J_i,\mu_i,L}(r)$ as a function of radial coordinate $r$ through the 3D transition density $\rho^{\alpha_f}_{\alpha_i}(\br)$
\beqn
\label{Definition_TD}
\lefteqn{
\langle \alpha_f| \hat\rho(\br) \, Y_{LM} | \alpha_i\rangle
}
\nonumber\\
 &=&\dfrac{(-1)^{2L}}{\hat J_f} \langle J_iM_i LM | J_fM_f\rangle
 \langle J_f\mu_f || \hat\rho(\br) Y_{L} || J_i \mu_i\rangle
\, .
      \nonumber \\
\eeqn
The left-hand-side of Eq.~(\ref{Definition_TD}) is given by
\beq
\label{Definition_TD2}
\langle \alpha_f| \hat\rho(\br) Y_{LM} | \alpha_i\rangle
= \int \! d\hat\br \;  \rho^{\alpha_f}_{\alpha_i}(\br) \, Y_{LM}(\hat\br)
\, .
\eeq
The reduced transition density $\rho^{J_f\mu_f}_{J_i\mu_i,L}(r)$ with angular momentum transfer $L$ is therefore given by
\beq
\label{Definition_TD3}
\rho^{J_f\mu_f}_{J_i\mu_i,L}(r)
= \hat J^{-1}_i \langle J_f\mu_f|| \hat{\rho}(\br) \, Y_{L} || J_i \mu_i\rangle
\, ,
\eeq
where the factor $\hat J^{-1}_i$ is introduced such that the integration of $r^{L+2}\rho^{J_f,\mu_f}_{J_i,\mu_i,L}(r)$ over the radial coordinate $r$ gives the value of the transition matrix element of multipolarity $L$, cf.\ Eq.~(\ref{TME}).

In terms of the reduced transition density, the longitudinal form factor $F_L(q)$ for angular momentum transfer
$L$ in Eq.~\eqref{FF} has the form
\begin{equation}
F_L (q)
= \dfrac{\sqrt{4\pi}}{Z}
   \int^\infty_0 \! dr \; r^2 \,  \rho^{J_f\mu_f}_{J_i\mu_i,L}(r) \, j_L (qr) \, .
\end{equation}
We note that our convention for the reduced transition density differs from
the one of Eq.~(5) of Ref.~\cite{Heisenberg82} by a factor of $\sqrt{4\pi}/Z$.

According to the asymptotic behavior of the spherical Bessel function
$j_L (qr)$ \cite{Uberall71}
\beqn
\lim_{qr \rightarrow 0}
j_L (qr)
& = & \dfrac{(qr)^L}{(2L+1)!!}
      \bigg[1 - \dfrac{1}{L+3/2} \bigg(\dfrac{qr}{2}\bigg)^2
      \nonumber \\
&   & \quad   + \dfrac{1}{2(L+3/2)(L+5/2)} \bigg(\dfrac{qr}{2}\bigg)^4
              - \ldots \bigg] \, ,
      \nonumber \\
\eeqn
the Coulomb form factor of inelastic scattering in the $q \rightarrow 0$
limit is given by~\cite{Rosen67,Uberall71}
\beqn
\label{FL}
\lefteqn{
F_L(q)
} \nonumber \\
& =& \dfrac{\sqrt{4\pi}}{Z} \dfrac{q^L}{(2L+1)!!}\sqrt{B(EL)}\nonumber\\
&&\times\left[1 - \dfrac{q^2 \, R^2_{\rm tr}}{2(2L+3)}
                + \dfrac{q^4 \, R^4_{\rm tr}}{8(2L+3)(2L+5)} -\ldots\right]
          \, ,
\nonumber \\
\eeqn
where the effective transition radii $R^n_{\rm tr}$, $(n=2$, $4)$, are
defined as
\beq
 R^n_{\rm tr}
 =\dfrac{\int \! dr \; r^{L+n+2} \,  \rho^{J_f\mu_f}_{J_i\mu_i,L}}
        {\int \! dr \; r^{L+2}   \,  \rho^{J_f\mu_f}_{J_i\mu_i,L}} \, .
 \eeq
From these properties, one can extract the multipolarity $L$
of the transition, the transition strength $B(EL)$, and the transition
radius $R^2_{\rm tr}$ from the data for the Coulomb form factor in the low-$q$
region. Usually, one introduces a $q$-dependent multipole transition
matrix element $M^p_L(q)$ for graphical comparisons of matrix elements
and Coulomb form factors at small $q$ values
\begin{equation}
\label{MLq}
M_L(q^2) = \dfrac{Z}{\sqrt{4\pi}} \dfrac{(2L+1)!!}{q^L} F^{\ell}_L(q) \, .
\end{equation}
For elastic scattering, $L=0$, $\alpha_f=\alpha_i=\alpha$ and the Coulomb form factor
becomes in the $q \rightarrow 0$ limit
 \beqn
 \label{F0}
 F_0(q)
 &=&\dfrac{\sqrt{4\pi}}{Z} \int^\infty_0 \! \! dr \, r^2 \; \rho^{J\mu}_{J\mu,0}(r) \, \dfrac{\sin(qr)}{qr} \nonumber\\
 &=& 1 - \dfrac{q^2}{3!} r_{\rm ch} + \ldots \, ,
\eeqn
where $r_{\rm ch}$ is the rms charge radius of the state $| J\mu\rangle$.


\section{Derivation of transition density between GCM states}
\label{append2}

In this section, we derive the form of the transition density between two arbitrary GCM states for the general case of triaxially deformed nuclei. In this case, the wave function of  GCM state is given by
\beqn
| \alpha \rangle = \sum_{K,q} F^{JK}_{\mu,q} \, \hat{P}^J_{MK} \, \hat{P}^N \, \hat{P}^Z | q \rangle \, .
\eeqn
Sandwiching the density operator $\hat \rho(\br)\equiv\sum_{i}\delta(\br-\br_i)$ between the wave functions of the initial $|\alpha_i\rangle$ and final $|\alpha_f\rangle$ GCM states, one obtains the 3D transition density $\rho^{\alpha_f}_{\alpha_i}(\br)$
\beq
\label{3DTD}
\rho^{\alpha_f}_{\alpha_i}(\br)
= \sum_{K_f,K_i}\sum_{q_f,q_i}
  F^{J_fK_f\ast}_{\mu_f,q'} F^{J_iK_i}_{\mu_i,q}  \rho^{\sigma_fq'}_{\sigma_iq}(\br) \, ,
\eeq
where we have introduced the shorthand notation $\sigma\equiv\{JMK\}$. The kernel of the 3D transition density $\rho^{\sigma_fq'}_{\sigma_iq}(\br)$ reads
\beqn
\lefteqn{
 \rho^{\sigma_fq'}_{\sigma_iq}(\br)
} \nonumber \\
& = & \dfrac{\hat J^2_i \, \hat J^2_f}{(8\pi^2)^2}
       \iint \! d\Omega^\prime \, d\Omega \;
       D^{J_f \ast}_{K_f M_f}(\Omega^\prime) \
       D^{J_i}_{K_iM_i}(\Omega)   \nonumber\\
  &&   \times \langle  q' | \hat R(\Omega^\prime) \, \hat{\rho}(\br) \,
       \hat{R}^\dagger(\Omega^\prime) \, \hat{P}^N \hat{P}^Z
       \hat{R}(\Omega^\prime) \ \hat{R}^\dagger(\Omega) | q \rangle
\, .\nonumber\\
\eeqn
For any HFB state $| q \rangle$, one has
\beqn
\langle q' | \hat R(\Omega^\prime) \hat\rho(\br)  \hat{R}^\dagger(\Omega^\prime) |  q   \rangle
&\equiv& \langle q' |\hat\rho(\tilde\br_{\Omega'}) | q  \rangle\nonumber\\
&=& \hat R^\dagger(\Omega^\prime)
  \big[\langle q' |\hat\rho(\br) | q  \rangle \big] \, ,
\eeqn
where $\tilde\br_{\Omega'}=D(\Omega^\prime) \, \br$.  Decomposing the rotation operator $\hat{R}(\Omega)\equiv\hat{R}(\Omega^{\prime\prime}) \, \hat{R}(\Omega^\prime)$, $\hat{R}^\dagger(\Omega)=\hat{R}^\dagger(\Omega^\prime)\hat R^\dagger(\Omega^{\prime\prime})$ and using the properties of Wigner $D$-functions
 \beq
 D^{J_i}_{K_iM_i}(\Omega)
 = \sum_{K} D^{J_i}_{K_iK}(\Omega^{\prime\prime}) \,
              D^{J_i}_{KM_i}(\Omega^\prime) \, ,
 \eeq
the kernel $\rho^{\sigma_fq'}_{\sigma_iq}(\br)$ of the 3D transition density in (\ref{3DTD}) can be simplified to
 \beqn
\label{transition-dens}
\rho^{\sigma_fq'}_{\sigma_iq}(\br)
 &=&
       \dfrac{\hat J^2_f}{8\pi^2}
       \int \! d\Omega^\prime \;  D^{J_f \ast}_{K_f M_f}(\Omega^\prime) \nonumber\\
   && \times
       \sum_{K} D^{J_i}_{KM_i}(\Omega^\prime) \,
       \hat{R}^\dagger(\Omega^\prime) \, \rho^{J_iKK_i}_{q'q}(\br) \, ,
\eeqn
where the $\rho^{J_iKK_i}_{q'q}(\br)$ is defined as
\beqn
\label{rhoJKK}
\rho^{J_iKK_i}_{q'q}(\br)  \equiv  \langle q'  |   \hat \rho(\br)
      \hat P^{J_i}_{KK_i}\hat{P}^N \hat{P}^Z | q  \rangle \, .
\eeqn


\section{Expansion in terms of spherical harmonics}
\label{append3}

To separate the radial dependence of the 3D transition density from its
trivial angular part, inspired by Ref.~\cite{Zaringhalam77} we expand
$\rho^{J_iKK_i}_{q'q}(\br)$ in
Eq.~(\ref{rhoJKK}) in terms of spherical harmonics
 \beq
 \rho^{J_iKK_i}_{q'q}(\br)
  = \sum^{\infty}_{\lambda=0}
    \sum^{\lambda}_{\nu=-\lambda} \rho^{J_iKK_i}_{q'q;\lambda\nu} (r) \,
    Y_{\lambda\nu}(\hat \br) \, ,
 \eeq
where the radial part $\rho^{J_i K K_i}_{q'q;\lambda\nu} (r)$ is given by
 \beq
 \label{rhoJKK_r}
 \rho^{J_iKK_i}_{q'q;\lambda\nu} (r)
 = \int \! d\hat\br \; \rho^{J_iKK_i}_{q'q}(r, \hat \br) \,
    Y^\ast_{\lambda\nu}(\hat \br) \, .
 \eeq
In this case, the rotation $\hat R^\dagger(\Omega^\prime)$ of $\rho^{J_iKK_i}_{q'q}(\br)$ in Eq.~(\ref{transition-dens}) can be evaluated
analytically
 \beq
 \hat R^\dagger(\Omega^\prime)  \rho^{J_iKK_i}_{q'q}(\br)
 =  \sum_{\lambda\nu\nu^\prime} D^{\lambda\ast}_{\nu\nu^\prime}(\Omega^\prime) \,
       \rho^{J_iKK_i}_{q'q;\lambda\nu} (r) \,
       Y_{\lambda\nu^\prime}(\hat \br) \, .
 \eeq
The kernel $\rho^{\sigma_fq'}_{\sigma_iq}(\br)$ of the 3D transition density in Eq.~(\ref{transition-dens}) becomes
 \beqn
\rho^{\sigma_fq'}_{\sigma_iq}(\br)
& = &
       \dfrac{\hat J^2_f}{\hat J^2_i}\sum_{K\lambda\nu\nu^\prime}
      \langle J_fK_f \lambda \nu | J_iK\rangle
      \langle  J_f M_f \lambda \nu^\prime | J_i M_i \rangle \nonumber\\
 &&\times  \rho^{J_iKK_i}_{q'q;\lambda\nu} (r) \,
       Y_{\lambda\nu^\prime}(\hat \br) \, ,
\eeqn
where we have expressed the integration of the product of three Wigner $D$-functions over Euler angles as the product of two Clebsch-Gordan coefficients, making the assumption that $J_i+J_f+\lambda$ is integer~\cite{Varshalovich88}
\beqn
\lefteqn{
\int d\Omega^\prime  D^{J_f \ast}_{K_f M_f}(\Omega^\prime)
       D^{J_i}_{KM_i}(\Omega^\prime) D^{\lambda\ast}_{\nu\nu^\prime}(\Omega^\prime)
} \nonumber\\
& = & \dfrac{8\pi^2}{\hat J^2_i}
      \langle J_f K_f \lambda \nu | J_i K\rangle \,
      \langle  J_f M_f \lambda \nu^\prime | J_i M_i \rangle \, .
\eeqn
By substituting the expression for $ \rho^{\alpha_f}_{\alpha_i}(\br)$ into Eqs.~(\ref{Definition_TD}), (\ref{Definition_TD2}), and (\ref{Definition_TD3}), one finds as an expression for the reduced transition density
\beqn
 \rho^{J_f \mu_f}_{J_i \mu_i,L}(r)
 &=& (-1)^{2L} \dfrac{\hat J^3_f}{\hat J^3_i}
 \sum_{K_f,K_i}\sum_{q'q}
      F^{J_fK_f\ast}_{\mu_f,q'} F^{J_iK_i}_{\mu_i,q}\nonumber\\
  &&  \times  \sum_{K\lambda\nu\nu^\prime}
      \langle J_fK_f \lambda \nu | J_iK\rangle
      \dfrac{\langle  J_f M_f \lambda \nu^\prime | J_i M_i \rangle}{\langle J_iM_i LM| J_fM_f\rangle} \nonumber\\
 &&\times  \rho^{J_iKK_i}_{q'q;\lambda\nu} (r)
       \int \! d\hat\br \; Y_{LM}(\hat{\br}) \, Y_{\lambda\nu^\prime}(\hat \br)
 \, .
\eeqn
With the help of the orthogonality relation of the spherical harmonics, $\int \! d\hat{\br} \; Y_{LM}(\hat\br) \, Y_{\lambda\nu^\prime}(\hat \br)=(-1)^{-M} \, \delta_{L\lambda} \; \delta_{M-\nu^\prime}$, and the symmetry relation
 \mbox{$\langle  J_f M_f L-M | J_i M_i \rangle$} $=(-1)^{2L-M+J_i-J_f}\dfrac{\hat J_i}{\hat J_f}\langle J_iM_i LM| J_fM_f\rangle$ of the Clebsch-Gordan coefficients, the reduced transition density can be simplified to
 \beqn
 \rho^{J_f \mu_f}_{J_i \mu_i,L}(r)
 &=& (-1)^{J_i-J_f} \dfrac{\hat J^2_f}{\hat J^2_i}
 \sum_{K_f,K_i}\sum_{q'q}
      F^{J_fK_f\ast}_{\mu_f,q'} F^{J_iK_i}_{\mu_i,q}\nonumber\\
  &&  \times
      \sum_{K\nu}
      \langle J_fK_f  L \nu | J_iK\rangle \, \rho^{J_iKK_i}_{q'q;L\nu} (r) \, ,
\eeqn
where we have replaced the phase factor $(-1)^{4L-2M+J_i-J_f}$ by $(-1)^{J_i-J_f}$. Substituting Eq.~(\ref{rhoJKK_r}) into the above equation, one finds as the final expression for the reduced transition density of triaxially deformed
nuclei
 \beqn
 \label{reducedTD}
\lefteqn{
 \rho^{J_f \mu_f}_{J_i \mu_i,L}(r)
} \nonumber \\
 &=& (-1)^{J_i-J_f}\dfrac{\hat J^2_f}{\hat J^2_i}
 \sum_{K_i,K_f}\sum_{q'q}
      F^{J_fK_f\ast}_{\mu_f,q'} F^{J_iK_i}_{\mu_i,q}\nonumber\\
  &&  \times
      \sum_{K\nu} \langle J_fK_f L \nu | J_iK\rangle
      \int \! d\hat{\br} \; \rho^{J_iKK_i}_{q'q}(\br) \, Y^\ast_{L\nu}(\hat{\br}) \, .
\nonumber\\
\eeqn

When axial symmetry about the $z$ axis is imposed on the intrinsic states
$| q\rangle$, all components with $K_i \neq 0$ and $K_f \neq 0$ vanish. In this case, the reduced transition density $\rho^{J_f \mu_f}_{J_i \mu_i,L}(r)$ in Eq.~(\ref{reducedTD}) is simplified as
 \beqn
 \label{reducedTD_axial}
 \rho^{J_f \mu_f}_{J_i \mu_i,L}(r)
 &=& (-1)^{J_i-J_f}\dfrac{\hat J^2_f}{\hat J^2_i} \sum_{q'q}
      F^{J_f0\ast}_{\mu_f,q'} F^{J_i0}_{\mu_i,q }\nonumber\\
&&\times
      \sum_{K} \langle J_f0 L K | J_i K\rangle
      \int \! d\hat{\br} \; \rho^{J_iK0}_{q'q}(\br) \, Y^\ast_{LK}(\hat{\br})
    \nonumber\\
&=& (-1)^{J_i-J_f} \dfrac{\hat J^2_f}{\hat J^2_i}
      \sum_{K} \langle J_f0 L K | J_i K\rangle \nonumber\\
&&\times
      \int \! d\hat{\br} \;  \rho^{J_fJ_iK0}_{\mu_f\mu_i}(\br) \,  Y^\ast_{LK}(\hat{\br}) 
\eeqn
where the \textit{pseudo GCM density} $\rho^{J_fJ_iK0}_{\mu_f\mu_i}(\br)$ has been defined in Eq.~(\ref{pseudo_density}).

\section{Multipole transition matrix elements}
\label{append4}
With the reduced transition density $\rho^{J_f \mu_f}_{J_i \mu_i,L}(r)$ (\ref{reducedTD}), one can calculate the multipole ($L$) transition matrix element straightforwardly
\beqn
 M^{J_f \mu_f}_{J_i \mu_i,L}
 &\equiv& \int \! dr \, r^{2+L} \; \rho^{J_f \mu_f}_{J_i \mu_i,L}(r) \nonumber\\
    &=&(-1)^{J_i-J_f}\dfrac{\hat J^2_f}{\hat J^2_i} \sum_{K_f,K_i}\sum_{q'q}
      F^{J_fK_f\ast}_{\mu_f,q'} F^{J_iK_i}_{\mu_i,q } \nonumber\\
 &&\times \sum_{K\nu} \langle J_fK_f L -\nu | J_iK\rangle(-1)^{-\nu}\nonumber\\
 &&\times
      \int \! d^3\br \; r^L \, Y_{L\nu}(\hat \br) \,
      \langle q' | \hat{\rho}(\br) \hat{P}^{J_i}_{KK_i} \hat{P}^N
      \hat{P}^Z | q  \rangle \, .
\nonumber\\
\eeqn
By defining the transition operator of multipolarity $L$ as $\hat Q_{L\nu}=r^L Y_{L\nu}$, and  using the relation between Clebsch-Gordan coefficients and $3j$-symbols~\cite{Varshalovich88}, we obtain the final expression for the multipole transition matrix element, 
\beqn
\label{TME2}
 M^{J_f \mu_f}_{J_i \mu_i,L}
&=&(-1)^{2J_i} \dfrac{\hat J^2_f}{\hat J_i} \sum_{K_f,K_i}\sum_{q'q}
      F^{J_fK_f\ast}_{\mu_f,q'} F^{J_iK_i}_{\mu_i,q}\nonumber\\
    &&\times
 \sum_{K\nu}(-1)^{J_f-K_f+2K}
   \begin{pmatrix}
  J_f & L &  J_i \\
  -K_f  &  \nu &  K
 \end{pmatrix}  \nonumber\\
    &&\times\langle q' |  \hat Q_{L\nu}
      \hat P^{J_i}_{KK_i}\hat{P}^N \hat{P}^Z | q  \rangle \, .
\eeqn
It can be easily shown that the electric multipole transition strength is given by
\begin{equation}
\label{BEL}
B(EL :J_i\mu_i \to J_f\mu_f)
= \Big| M^{J_f \mu_f,p}_{J_i \mu_i,L} \Big|^2 \, ,
\end{equation}
provided that the operator $\hat Q_{L\nu}$ is replaced by the electric one $\hat Q_{L\nu}=er^LY_{L\nu}$.


%

\end{appendix}



\begin{thebibliography}{99}

\bibitem{Hofstadter56}
R. Hofstadter, Rev. Mod. Phys. \textbf{28}, 214 (1956).

\bibitem{Alder56}
K. Alder, {\AA}. Bohr, T. Huus, B. Mottelson, and A. Winther,
Rev. Mod. Phys. \textbf{28}, 432 (1956).

\bibitem{Forest66}
T. de Forest, Jr. and J. D. Walecka,
Adv. Phys. \textbf{15}, 1 (1966).

\bibitem{Uberall71}
H. {\"U}berall,
\textit{Electron Scattering from Complex Nuclei},
Parts A and B,
Academic Press, New York, 1971.

\bibitem{Barrett74}
R. C. Barrett,
Rep. Prog. Phys. \textbf{37}, 1 (1974).

\bibitem{Dreher74a}
B. Dreher, J. Friedrich, K. Merle, H. Rothhaas, G. L{\"u}hrs,
Nucl. Phys. \textbf{A235}, 219 (1974).

\bibitem{Donnelly75}
T. W. Donnelly and J. D. Walecka,
Annu. Rev. Nucl. Part. Sci. \textbf{25}, 329 (1975).

\bibitem{Friar75}
J. L. Friar and J.W. Negele,
Adv. in Nucl. Phys. \textbf{8}, 219 (1975).

\bibitem{Heisenberg81}
J. Heisenberg,
Adv. Nucl. Phys. \textbf{12}, 61 (1981).

\bibitem{Heisenberg83}
J. Heisenberg and H. P. Blok,
Ann. Rev. Nucl. Part. Sci. \textbf{33}, 569 (1983).

\bibitem{Donnelly84}
T. W. Donnelly and I. Sick,
Rev. Mod. Phys. \textbf{56}, 461 (1984).

\bibitem{Sick85}
I. Sick,
in \textit{Advanced Methods in the Evaluation of Nuclear
Scattering Data},
Lecture Notes in Physics Vol. \textbf{236}, 137 (1985).

\bibitem{Vries87}
H. de Vries, C. W. de Jager, and C. de Vries,
At. Data Nucl. Data Tables 36, 495 (1987).

\bibitem{Frois87}
B. Frois and C. N. Papanicolas,
Ann. Rev. Nucl. Part. Sci. \textbf{37}, 133 (1987).

\bibitem{Hodgson92}
P. E. Hodgson,
Hyperfine Interactions \textbf{74}, 75 (1992).

\bibitem{Walecka04a}
J. D. Walecka,
\textit{Electron Scattering for Nuclear and Nucleon Structure},
(Cambridge University Press, Cambridge, 2004).

\bibitem{FV82}
J. Friedrich and N. Voegler,
Nucl. Phys. \textbf{A459}, 192 (1982).

\bibitem{Friedrich86}
J. Friedrich, N. Voegler, and P.-G. Reinhard,
Nucl. Phys. \textbf{A459}, 10 (1986).

\bibitem{Wakasugi04}
M. Wakasugi, T. Suda, and Y. Yano,
Nucl. Instrum. Methods Phys. Res., Sect. \textbf{A532}, 216 (2004).

\bibitem{Suda05}
T. Suda and M. Wakasugi,
Prog. Part. Nucl. Phys. \textbf{55}, 417 (2005).

\bibitem{Suda09}
T. Suda, M. Wakasugi, T. Emoto, K. Ishii, S. Ito, K. Kurita,
A. Kuwajima, A. Noda, T. Shirai, T. Tamae, H. Tongu, S. Wang,
and Y. Yano,
Phys. Rev. Lett. \textbf{102}, 102501 (2009).

\bibitem{Simon07a}
H. Simon,
Nucl. Phys. \textbf{A787}, 102c (2007).

\bibitem{Antonov11}
A. N. Antonov \textit{et al.},
Nucl. Instrum. Methods Phys. Res., Sect. \textbf{A637}, 60 (2011).

\bibitem{Helm56}
R. H. Helm,
Phys. Rev. \textbf{104}, 1466 (1956).

\bibitem{Tassie56}
L. J. Tassie,
Aust. J. Phys. \textbf{9}, 407 (1956).

\bibitem{Brown83}
B. A. Brown, R. Radhi, and B. H. Wildenthal,
Phys. Rep. \textbf{101},  313 (1983).

\bibitem{Dieperink78a}
A. E. L. Dieperink, F. Iachello, A. Rinat, and C. Creswell,
Phys. Lett. \textbf{B76}, 135 (1978);
A. E. L. Dieperink, Nucl. Phys. \textbf{A358}, 189c (1981).

\bibitem{Horikawa77}
Y. Horikawa, T. Hoshino, and A. Arima,
Nucl. Phys. \textbf{A278}, 297 (1977).

\bibitem{Sagawa87}
H. Sagawa, O. Scholten, and B. A. Brown,
Nucl. Phys. \textbf{A462}, 1 (1987).

\bibitem{Yokoyama89}
A. Yokoyama and K. Ogawa,
Phys. Rev. C \textbf{39}, 2458 (1989).

\bibitem{Radhi03}
R. A. Radhi and A. Bouchebak,
Nucl. Phys. \textbf{A716}, 87 (2003).

\bibitem{Karataglidis07}
S. Karataglidis and K. Amos,
Phys. Lett. \textbf{B650}, 148 (2007).

\bibitem{Radhi08a}
R. A. Radhi, A. A. Abdullah, and A. H. Raheem,
Nucl. Phys. \textbf{798}, 16 (2008).

\bibitem{Bender03}
M. Bender, P.-H. Heenen, and P.-G. Reinhard,
Rev. Mod. Phys. \textbf{75}, 121 (2003).

\bibitem{Negele70}
J. W. Negele,
Phys. Rev. C \textbf{1}, 1260 (1970).

\bibitem{Miller72}
L. D. Miller and A. E. S. Green,
Phys. Rev. C \textbf{5}, 241 (1972).

\bibitem{Vautherin72}
D. Vautherin and D. M. Brink,
Phys. Rev. C \textbf{5}, 626 (1972).

\bibitem{Decharge80}
J. Decharg{\'e} and D. Gogny,
Phys. Rev. C \textbf{21}, 1568 (1980).

\bibitem{Wang04}
Z. Wang and Z. Ren,
Phys. Rev. C \textbf{70}, 034303 (2004).

\bibitem{Antonov05}
A. N. Antonov, D. N. Kadrev, M. K. Gaidarov, E. Moya de Guerra,
P. Sarriguren, J. M. Udias, V. K. Lukyanov, E. V. Zemlyanaya, and
G. Z. Krumova,
Phys. Rev. C \textbf{72}, 044307 (2005).

\bibitem{Roca08a}
X. Roca-Maza, M. Centelles, F. Salvat, and X. Vi{\~n}as,
Phys. Rev. C \textbf{78}, 044332 (2008).

\bibitem{Roca12a}
X. Roca-Maza, M. Centelles, F. Salvat, and X. Vi{\~n}as
Phys. Rev. C \textbf{87}, 014304 (2012).

\bibitem{Negele77}
J. W. Negele and G. Rinker,
Phys. Rev. C \textbf{15}, 1499 (1977).

\bibitem{Moya80a}
E. Moya de Guerra,
Ann. Phys. (NY) \textbf{128}, 286 (1980).

\bibitem{Sarriguren89}
P. Sarriguren, E. Graca, D. W. L. Sprung, E. Moya de Guerra,
and D. Berdichevsky,
Phys. Rev. C \textbf{40}, 1414 (1989).

\bibitem{Berdichevsky88}
D. Berdichevsky, P. Sarriguren, E. Moya de Guerra, M. Nishimura,
and D. W. L. Sprung,
Phys. Rev. C \textbf{38}, 338 (1988).

\bibitem{Faessler76}
A. Faessler, S. Krewald, A. Plastino and J. Speth,
Z. Phys. \textbf{A 276}, 91 (1976).

\bibitem{Reinhard79}
P.-G. Reinhard and S. Drechsel,
Z. Phys. \textbf{A 290}, 85 (1979).

\bibitem{Gogny79}
D. Gogny,
in \textit{Nuclear Physics with Electromagnetic Interactions},
H. Arenh{\"o}vel and D. Drechsel [edts.],
Lecture Notes in Physics, Vol. 108 (Springer-Verlag, New York, 1979), p. 88.

\bibitem{Decharge83}
J. Decharg{\'e}, M. Girod, D. Gogny and B. Grammaticos,
Nucl. Phys. \textbf{A358}, 203c (1983).

\bibitem{Esbensen83}
H. Esbensen and G. F. Bertsch,
Phys. Rev. C \textbf{28}, 355 (1983).

\bibitem{Barranco87}
F. Barranco and R. A. Broglia,
Phys. Rev. Lett. \textbf{59}, 2724 (1987).

\bibitem{Johnson88}
M. B. Johnson and G. Wenes,
Phys. Rev. C \textbf{38}, 386 (1988).

\bibitem{Sil08}
T. Sil and S. Shlomo,
Phys. Scr. \textbf{78}, 065202 (2008).

\bibitem{Nobre11}
G. P. A. Nobre, F. S. Dietrich, J. E. Escher, I. J. Thompson,
M. Dupuis, J. Terasaki, and J. Engel,
Phys. Rev. C \textbf{84}, 064609 (2011).

\bibitem{Abgrall74a}
Y. Abgrall, P. Gabinski, and J. Labarsouque,
Nucl. Phys. \textbf{A232}, 235 (1974).

\bibitem{Zaringhalam77}
Z. Zaringhalam and J. W. Negele,
Nucl. Phys. \textbf{A288}, 417 (1977).

\bibitem{Moya78a}
E. Moya de Guerra and A. E. L. Dieperink,
Phys. Rev. C \textbf{18}, 1596 (1978).

\bibitem{Guerra80}
E. Moya de Guerra and S. Kowalski,
Phys. Rev. C \textbf{20}, 357 (1979);
Phys. Rev. C \textbf{22}, 1308 (1980).

\bibitem{Dieperink87}
A. E. L. Dieperink and E. Moya de Guerra,
Phys. Lett. \textbf{B189}, 267 (1987).

\bibitem{Graca88}
E. Graca, P. Sarriguren, D. Berdichevsky, D. W. L. Sprung,
E. Moya De Guerra, M. Nishimura,
Nucl. Phys. \textbf{A483}, 77 (1988).

\bibitem{Nishimura88}
M. Nishimura, D. W. L. Sprung, and E. Moya De Guerra,
Phys. Lett. \textbf{B161}, 235 (1985).

\bibitem{Guerra86}
E. Moya de Guerra,
Phys. Rep. \textbf{138}, 293 (1986).

\bibitem{Yao12}
J. M. Yao, S. Baroni, M. Bender, and P.-H. Heenen,
Phys. Rev. C \textbf{86}, 014310 (2012).

\bibitem{Yao13a}
J. M. Yao, H. Mei, and Z. P. Li,
Phys. Lett. \textbf{B723}, 459 (2013).

\bibitem{Wu14a}
X. Y. Wu, J. M. Yao, and Z. P. Li,
Phys. Rev. C \textbf{89}, 017304 (2014).

\bibitem{Mei14}
H. Mei, K. Hagino, J. M. Yao, and T. Motoba,
Phys. Rev. C \textbf{90}, 064302 (2014).

\bibitem{Fuk13}
Y. Fukuoka, S. Shinohara, Y. Funaki, T. Nakatsukasa, and K. Yabana,
Phys. Rev. C \textbf{88}, 014321 (2013).

\bibitem{Ter96a}
J. Terasaki, P.-H. Heenen, H. Flocard, and P. Bonche,
Nucl. Phys. \textbf{A600}, 371 (1996).

\bibitem{Bonche05}
P. Bonche, H. Flocard, and P.-H. Heenen,
Comput. Phys. Comm. \textbf{171}, 49 (2005).

 \bibitem{Chabanat98}
E. Chabanat, P. Bonche, P. Haensel, J. Meyer, and R. Schaeffer,
Nucl. Phys. \textbf{A635}, 231 (1998);
Nucl. Phys. \textbf{A643}, 441(E) (1998).

\bibitem{Rigollet99}
C. Rigollet, P. Bonche, H. Flocard, and P.-H. Heenen,
Phys. Rev. C \textbf{59}, 3120 (1999).

\bibitem{Ring80}
P. Ring and P. Schuck,
\textit{The Nuclear Many-Body Problem}
(Springer, Heidelberg, 1980).

\bibitem{Lacroix09}
D. Lacroix, T. Duguet, and M. Bender,
Phys. Rev. C \textbf{79}, 044318 (2009).

\bibitem{Bender08}
M. Bender and P.-H. Heenen,
Phys. Rev. C \textbf{78}, 024309 (2008).

\bibitem{Heisenberg82}
J. Heisenberg, J. Lichtenstadt, C. N. Papanicolas, and J. S. McCarthy,
Phys. Rev. C \textbf{25}, 2292 (1982).

\bibitem{Angeli04}
I. Angeli,
At. Data Nucl. Data Tables \textbf{87}, 185 (2004).

\bibitem{Schmid90a}
K. W. Schmid and F. Gr{\"u}mmer,
Z. Phys. \textbf{A337}, 267 (1990).

\bibitem{Schmid91a}
K. W. Schmid and  P.-G. Reinhard,
Nucl. Phys. \textbf{A530}, 283 (1991).

\bibitem{Rodriguez04a}
R. R. Rodr{\'i}guez-Guzm{\'a}n and K. W. Schmid,
Eur. Phys. J. \textbf{A 19}, 45 (2004).

\bibitem{Tassie58}
L. J. Tassie and F. C. Barker,
Phys. Rev. \textbf{111}, 940 (1958).

\bibitem{Yao10}
J. M. Yao, J. Meng, P. Ring and D. Vretenar,
Phys. Rev. C \textbf{81}, 044311 (2010);
J. M. Yao, K. Hagino, Z. P. Li, J. Meng, and P. Ring,
Phys. Rev. C 89, 054306 (2014).

\bibitem{Rodriguez10}
T. R. Rodr\'{\i}guez and J. L. Egido,
Phys. Rev. C \textbf{81}, 064323 (2010).

\bibitem{Baye86}
D. Baye and P.-H. Heenen,
J. Phys. \textbf{A19}, 2041 (1986).

\bibitem{Valor00}
A. Valor, P. H. Heenen, and P. Bonche,
Nucl. Phys. \textbf{A671}, 145 (2000).

\bibitem{Terrien73}
Y. Terrien,
Nucl. Phys. \textbf{A199}, 65 (1973);
Nucl. Phys. \textbf{A215}, 29 (1973).

\bibitem{Guzman02}
R. Rodr\'{\i}guez-Guzm\'{a}n, J. L. Egido, and L. M. Robledo,
Nucl. Phys. \textbf{A709}, 201 (2002).

\bibitem{Niksic06}
T. Nik\v{s}i\'{c}, D. Vretenar, and P. Ring,
Phys. Rev. C \textbf{73}, 034308 (2006);
Phys. Rev. C \textbf{74}, 064309 (2006).

\bibitem{Li74}
G. C. Li, M. R. Yearian, and I. Sick,
Phys. Rev. C \textbf{9}, 1861 (1974).

\bibitem{Johnston74}
A. Johnston and T. E. Drake,
J. Phys. A \textbf{7}, 898 (1974).

\bibitem{Zarek78}
H. Zarek, S. Yen, B. O. Pich, T. E. Drake, C. F. Williamson, S. Kowalski,
C. P. Sargent, W. Chung, B. H. Wildenthal, M. Harvey, and H. C. Lee,
Phys. Lett. \textbf{B80}, 26 (1978).

\bibitem{Orce08}
J. N. Orce, B. Crider, S. Mukhopadhyay, E. Peters, E. Elhami, M. Scheck,
B. Singh, M. T. McEllistrem and S. W. Yates,
Phys. Rev. C \textbf{77}, 064301 (2008).

\bibitem{Broda12}
R. Broda, T. Pawwat, W. Krlas, R. V. F. Janssens, S. Zhu, W. B. Walters,
B. Fornal, C. J. Chiara, M. P. Carpenter, N. Hoteling, W. Iskra,
F. G. Kondev, T. Lauritsen, D. Seweryniak, I. Stefanescu, X. Wang,
and J. Wrzesiski,
Phys. Rev. C \textbf{86}, 064312 (2012).

\bibitem{Allmond14}
J. M. Allmond, B. A. Brown, A. E. Stuchbery, A. Galindo-Uribarri,
E. Padilla-Rodal, D. C. Radford, J. C. Batchelder, M. E. Howard,
J. F. Liang, B. Manning, R. L. Varner, and C.-H. Yu,
Phys. Rev. C \textbf{90}, 034309 (2014).

\bibitem{Sim02}
R. F Simoes, D. S Monteiro, L. K Ono, A. M. Jacob, J. M. B Shorto,
N. Added, and E. Crema,
Phys. Lett. \textbf{B527}, 187 (2002).

\bibitem{Stefanini95}
A. M. Stefanini, D. Ackermann, L. Corradi, D. R. Napoli, C. Petrache,
P. Spolaore, P. Bednarczyk, H. Q. Zhang, S. Beghini, G. Montagnoli,
L. Mueller, F. Scarlassara, G. F. Segato, F. Soramel, and N. Rowley,
Phys. Rev. Lett. \textbf{74}, 864 (1995).

\bibitem{Garrett10a}
P. E. Garrett and J. L. Wood,
J. Phys. G \textbf{37}, 064028 (2010).

\bibitem{Garrett08a}
P. E. Garrett, K. L. Green, and J. L. Wood,
Phys. Rev. C \textbf{78}, 044307 (2008)

\bibitem{Ficenec70}
J. R. Ficenec, W. P. Trower, J. Heisenberg, and I. Sick,
Phys. Lett. \textbf{B32}, 460 (1970).

\bibitem{Sick75}
I. Sick, J. B. Bellicard, M. Bernheim, B. Frois, M. Huet, Ph. Leconte,
J. Mougey, Phan Xuan-Ho, D. Royer, and S. Turck,
Phys. Rev. Lett. \textbf{35}, 910 (1975).

\bibitem{Girod76}
M. Girod and D. Gogny,
Phys. Lett. \textbf{B64}, 5 (1976).

\bibitem{Pritychenko11}
B. Pritychenko, J. Choquette, M. Horoi, B. Karamy, and B. Singh,
At. Data Nucl. Data Tables \textbf{98}, 798 (2012).

\bibitem{Chaumeaux78}
A. Chaumeaux, V. Layly and R. Schaeffer,
Ann. Phys. (NY) \textbf{116}, 247 (1978).

\bibitem{Lombard81}
R. M. Lombard, G. D. Alkhazov, and O. A. Domchenkov,
Nucl. Phys. \textbf{A360}, 233  (1981).

\bibitem{Duguay67}
M. A. Duguay, C. K. Bockelman, T. H. Curtis, and R. A. Eisenstein,
Phys. Rev. \textbf{163}, 1259 (1967).

\bibitem{Torizuka69}
Y. Torizuka, Y. Kojima, M. Oyamada, K. Nakahara, K. Sugiyama, T. Terasawa,
K. Itoh, A. Yamaguchi, and M. Kimura,
Phys. Rev. \textbf{185}, 1499 (1969).

\bibitem{Frois83}
B. Frois, S. Turck-Chieze, J. B. Bellicard, M. Huet, P. Leconte, X.-H. Phan,
I. Sick, J. Heisenberg, M. Girod, K. Kumar, and B. Grammaticos,
Phys. Lett. \textbf{B122}, 347 (1983).

\bibitem{Braunstein88}
M. R. Braunstein, J. J. Kraushaar, R. P. Michel, J. H. Mitchell,
R. J. Peterson, H. P. Blok, and H. de Vries,
Phys. Rev. C \textbf{37}, 1870 (1988).

\bibitem{Bally14}
B. Bally, B. Avez, M. Bender, and P.-H. Heenen,
Phys. Rev. Lett. \textbf{113}, 162501 (2014).

\bibitem{BBH08}
M. Bender, G. F. Bertsch, and P.-H. Heenen,
Phys. Rev. C \textbf{78}, 054312 (2008).

\bibitem{Raphael70}
R. Raphael and M. Rose,
Phys. Rev. C \textbf{1}, 547 (1970).

\bibitem{Friar85}
J. L. Friar and W. C. Haxton,
Phys. Rev. C \textbf{31}, 2027 (1985).

\bibitem{Rosen67}
M. Rosen, R. Raphael, and H. {\"U}berall,
Phys. Rev. \textbf{163}, 927 (1967).

\bibitem{Varshalovich88}
D. A. Varshalovich, A. N. Moskalev and V. K. Khersonskii,
\textit{Quantum Theory of Angular Momentum} (World Scientific, 1988).

\end{thebibliography}
\end{document}